\documentclass[useAMS,usenatbib]{mn2e}
\usepackage{graphicx}
\usepackage{txfonts}
\usepackage{wrapfig}
\usepackage{longtable}
\usepackage[usenames]{color}
\def\aj{AJ}%
\def\apj{ApJ}%
\def\apjl{ApJ}%
\def\apjs{ApJS}%
\def\aap{A\&A}%
\def\aapr{A\&A~Rev.}%
\def\mnras{MNRAS}%
\begin{document}
\title[Bar in disk galaxies]{Quantifying the role of bars in the
build-up of central mass concentrations in disk galaxies}

\author[Jing ~Wang et al.]
       {Jing Wang$^{1,2}$\thanks{Email: wangj@mpa-garching.mpg.de},
       Guinevere Kauffmann$^2$, Roderik Overzier$^2$, Linda J. Tacconi$^3$,
       \newauthor Xu Kong$^{1,4}$, Amelie Saintonge$^{2,3}$, Barbara Catinella$^2$, David Schiminovich$^5$,
       \newauthor Sean M. Moran$^6$, Benjamin Johnson$^7$\\
       $^1$Center for Astrophysics, University of Science and Technology of China,
        230026 Hefei, China
       \\
       $^2$Max--Planck--Institut f\"ur Astrophysik,
        Karl--Schwarzschild--Str. 1, D-85741 Garching, Germany
        \\
       $^3$Max--Planck--Institut f\"ur extraterrestrische Physik,
       D-85741 Garching, Germany
        \\
       $^4$Key Laboratory for Research in Galaxies and Cosmology, 
   University of Science and Technology of China, Chinese Academy of 
   Sciences, China\\
       $^5$Department of Astronomy, Columbia University, New York, NY 10027, USA\\
       $^6$Department of Physics and Astronomy, The Johns Hopkins University, Baltimore, MD 21218, USA\\
       $^7$Institute of Astronomy, University of Cambridge, Cambridge CB3 0HA
}

\date{Accepted 2011 ???? ??
      Received 2011 ???? ??;
      in original form 2011 May}

\pubyear{2010}
\maketitle

\begin{abstract}
We analyze the role of bars in the build-up of
central mass concentrations in massive, disk galaxies.  Our parent sample consists
of  3757 face-on disk galaxies with redshifts between  0.01 and 0.05, selected from
the seventh Data Release of the Sloan Digital Sky Survey. 1555 galaxies with bars 
are identified using position angle and ellipticity profiles 
of the $i$-band light.  We compare the 
ratio of the specific star formation rate measured in the
1-3 kpc central region of the galaxy to
that measured for the whole galaxy.  Galaxies with strong bars 
have centrally enhanced star formation; 
the degree of enhancement depends primarily on the ellipticity of the bar, 
and not on the size of the bar or on  
the mass or  structure of the host galaxy.
The fraction of galaxies with
strong bars is highest at  stellar masses greater than $3 \times 10^{10}
M_{\odot}$, stellar surface densities less than $3 \times 10^8 M_{\odot}$
and concentration indices less than 2.5. In this region of parameter
space, galaxies with strong bars either have enhanced central star formation
rates, or star formation that is
{\em suppressed} compared to the mean.
This suggests that bars may play a role in the eventual quenching of  star
formation in galaxies.  Only 50\%  of galaxies with strongly concentrated
star formation have strong bars, indicating that other processes such as
galaxy interactions also induce central star-bursts.  We also find that the
ratio of the size of the bar to that of the disk depends mainly  on 
the colour of the galaxy, suggesting that the growth and destruction of  bars are  
regulated by gas accretion, as suggested by simulations.
\end{abstract}


\section{introduction}
It has long been conjectured that spiral galaxies evolve along the 
Hubble sequence from ``late-type'' disk-dominated systems
to ``early-type'' galaxies with very  massive bulges .  
This scenario is supported by the  fact that the
co-moving number density of massive bulge-dominated galaxies 
 increases from  high redshifts to 
the present day, while the number density of lower mass star-forming disk galaxies 
evolves comparatively little with redshift  \citep{Couch98, Fasano00, Kovac10}.  
It is believed that major merger events build 
classical bulges with light profiles that follow
a  $r^{1/4}$ law \citep{Barnes92,Toomre77},  
while pseudo-bulges
with light profiles that are close to
exponential are formed from the disk itself \citep[see reviews by][]{KK04,Wyse97}. 
Pseudo-bulges are  common in disk-dominated galaxies at low redshift  
\citep{Balcells03, Laurikainen07, Weinzirl09}. 

There are many mechanisms that may cause disks to become unstable and
gas to flow towards the center of the galaxy, eventually forming a bulge.
Tidal forces exerted by an interacting companion cause  
gas to lose angular momentum and to flow into the central region 
of the  galaxy where it  forms stars at an 
elevated rate (a so-called ``starburst'')\citep{Mihos96,DiMatteo07,Cox08}. 
 More than half of the galaxies with the highest central specific star formation
rates (SSFR)  in the local universe are interacting with 
a close companion \citep{Li08}. Alternatively, 
asymmetric structures in the disk, including bars and spiral arms 
exert torques that drive gas inwards  
\citep[see the review by][]{KK04}. 

The mechanisms by which bars influence the flow of gas in galaxies
have been studied in considerable detail using simulations. Bars induce  
gravitational torques that drive the gas towards the 
leading end of the bar where it is compressed and 
shocked, and where it loses energy by radiative 
processes, causing it to flow to the  center of the 
galaxy \citep {Ath92,Knapen02,Piner95,Regan04,Zurita08,Sheth02}. 
Galaxies with thick bars tend to form nuclear rings 
\citep{Regan03,Piner95,Schinnerer03}. Some of the gas 
will reach the center and  produce a disc-like bulge \citep{Ath92,Ath05}. 
Bars are robust if the galaxy remains undisturbed \citep{Ath05b}.

Observational evidence that this basic picture is correct
have come from highly resolved maps of the ionized, atomic and molecular gas 
in individual barred galaxies \citep[e.g. ][]{Quillen95, Schinnerer02, Sheth00},
which allow one to track the motion of the gas in detail. In addition,
there have been studies of samples of barred and unbarred galaxies that have
revealed  higher gas concentrations \citep{ Jogee05, Sakamoto99, Sheth05}, 
higher central star formation rates and 
flatter chemical abundance gradients 
\citep{deJong84,Devereux87,Ho97, Hawarden86, Martin94, Puxley88, Zaritsky94}
in the former. On average, these differences are 
most pronounced in galaxies with the strongest bars 
\citep{Ho97, Martin95, Martin94}. 
Recent observations suggest
that most disk galaxies at high redshift are extremely gas-rich
and hence dynamically unstable, and will likely
form bulges on a short timescale
(Genzel et al. 2006, 2008).  

It would be valuable  to quantify  the role that bars have played
in the formation of the present-day  bulge and pseudo-bulge population. 
It has been controversy about whether  bars are found more frequently in early-type or in late-type disk galaxies  \citep{Barazza08,Marinova09,Sheth08}.  Enhanced 
central star formation rates are more likely to be found in 
early-type galaxies  with bars than in late-type 
galaxies with bars \citep{Devereux87, Ho97}. 
These results are somewhat difficult to interpret because
Hubble type is a complex function of both bulge-to-disk ratio and
star formation rate.

In semi-analytic models of galaxy formation, disks form when gas cools
and settles at the center of a dark matter halo, while conserving angular 
momentum. In recent years,
simple prescriptions for bulge growth through disk instabilities have been
included in these models                              
\citep{Benson10,DeLucia11,Parry09}, but these have only been tested against
observations in a rudimentary way. Ideally, these models should be able to
account for the fraction of disk galaxies with pseudo-bulges as a function of
both halo mass and stellar mass and the size of the disk.   

In this paper, we study the effect of bars in building the central mass
concentrations in  galaxies as a function of parameters such as
stellar mass, stellar surface mass density and concentration. 
We make use of imaging data from the  Sloan Digital Sky Survey 
\citep[SDSS,][]{York00}. 
The SDSS images have a pixel scale of $0.''396$ and a
mean PSF (point spread function) 
of $1.''4$ (FWHM).  In this study, we identify  bars from the images of
$\sim$4000  disk galaxies with redshifts in the range 0.01 to 0.05 drawn 
from the seventh data release (DR7) of the SDSS. We first investigate the  
bar strength needed to cause enhancement in the  
central star formation rate of the galaxy. We then divide our galaxies into a
4-dimensional parameter 
space of stellar mass, concentration, stellar mass surface density and colour
and quantify the fraction of galaxies that are undergoing bar-driven central
mass growth at the present day. 

In Section 2, we describe our sample and the methods used to identify
bars  and to quantify their strength.
In Section 3.1, we examine how strongly star formation is
concentrated towards the centers  of barred galaxies  compared to control samples 
of non-barred galaxies matched in stellar mass, stellar mass 
surface density and colour. In this analysis, we parameterize
bar strength using  the ellipticity of the bar, $e_{bar}$.
 In Section 3.2 we examine the colours of bars, showing that
the strongest bars have  the bluest colours. In
Section 3.3, we show how the fraction of barred  galaxies 
varies according to the location of the galaxy in the 4-dimensional
parameter space of stellar mass, stellar mass surface density, concentration
and colour. In Section 3.4, we
investigate which galaxy properties correlate most strongly 
with bar strength and size. Finally, in
section 4, we summarize and discuss our results.

\section{Data}
\subsection{The Sample}

We select galaxies with M$_*>10^{10} M_{\odot}$ in the redshift 
range 0.01$<z<$0.05 from the MPA$\slash$JHU spectroscopic 
catalogue (http://www.mpa-garching.mpg.de/SDSS/DR7/),
which is drawn from the seventh data release (DR7) of the SDSS
\citep{Abazajian09}. This yields a sample of 16573 galaxies.  
Our aim is to assess the role of
bars in building bulges
in present-day disk-dominated systems, 
so we also select systems  with  $R_{90}/R_{50}<2.6$,
where R$_{90}$ and R$_{50}$ are the 
radii of the circular apertures enclosing  90 percent and 50 percent of the total
$r$-band light from the  galaxy. 
We further limit the sample to galaxies
with ellipticiy $1-b/a<0.25$ (or an inclination of 
less than 41.4 degrees), where $a$ and $b$ are the major and minor axes of an
ellipsoidal fit in the $r-$ band to each 
galaxy. We employ this criterion to select relatively face-on galaxies, 
because it is difficult to
identify bars in more 
inclined galaxies \citep{Laurikainen02}. 
These cuts yield a sample of 3890 galaxies,
which we will use to
characterize bars.

In this paper, we deribe global  star formation rates 
by fitting the observed  $FUV$, $NUV$,
$u$, $g$, $r$, $i$ and $z$ band fluxes of the galaxy
with a library of model SEDs \citep{Saintonge11, Wang11}. The model SED library is
generated using stellar population synthesis
models  \citep{Bruzual03} and includes model ``galaxies'' spanning a range in
metallicity, age, star formation history and dust 
attenuation strength. The probability that the observed SED can be described by each
model SED is calculated,
and the SFR of the galaxy is then given as the  probability-weighted  SFR averaged
over  the whole model library. 
We also use the $A_v$ derived using this  method to correct  the 
global colours of galaxies for internal reddening.  

We also make use of star formation rates 
estimated directly from the SDSS spectra, which are obtained through 3 arcsecond 
diamteter fibres. These star formation rates, which are
taken directly from the MPA/JHU catalogue,  are obtained by fitting the
measured emission line fluxes to a set of photo-ionization+stellar population
synthesis models \citep{Brinchmann04}. 

Finally, we derive morphological parameters, such as asymmetry index 
for all the galaxies \citep[see][for more details]{Wang11}.

\subsection {Bar Identification}

We use position angle and ellipticity profiles to identify bars in galaxies.
Similar methods have been employed in many past papers 
\citep[e.g. ][]{Barazza08, Jogee04, Knapen02, Laine02, Menendez07, Sheth03, Whyte02, Wozniak95}. 

We first use SExtractor to estimate galaxy  sizes (Petrosian radius,
R$_{50}$ and R$_{90}$), ellipticities and disk  position 
angles. We  mask  neighboring sources in the vicinity  of the main galaxy.  We then
run the $IRAF.ellipse$ 
task twice on  the SDSS $r-$ band images over a  range of radii extending   
from  twice the Petrosian radius down to a radius of  0.8 arcsec from the centre of
the galaxy.

The sampling radius in the fit increases exponentially with a step of 1.1, and the
center of each ellipse is 
allowed to vary.  During the first run of the $ellipse$ task, we set R$_{90}$ as the
starting radius. During the second run, 
we set the  starting radius to  max(R$_{50}$, 1.5'').  During
both runs, the initial values
of ellipticity and position angle are set equal to the values measured by 
SExtractor for the whole galaxy. 
The two profiles are then merged and  bad points are discarded. Bad points are
defined as  points where the $ellipse$ task failed to find a solution, 
the error on the ellipticity measurement is larger than 0.05, or the error on 
the position angle is larger than 5. 

The fit is said to have been successful if the merged profile covers the radius
ranging from 1.4 arcsec to R$_{90}$, and
if it  covers more than 60$\%$ of the original range in radii that served as input
for the $IRAF.ellipse$ task.  
We also visually examined all the fits and found that our procedure was successful in
almost all cases. Failures occurred for 133 galaxies (less than $\sim 3.5\%$ of the sample), 
mainly because the galaxies were  strongly disturbed 
or had extended low surface brightness disks. We note that \citet{Barazza08}
and \citet{Jogee04}
obtained a similar failure rate in their own analysis of bars in galaxies in the
Sloan Digital Sky Survey.
The remaining 3757 galaxies are well fitted with ellipses throughout the entire
disk, and our analysis of bars will be confined
to these systems from now on. We do not  deproject the images or the profiles before 
classifying a galaxy as barred or un-barred. Such de-projections are subject to
considerable
uncertainty, particularly for more bulge-dominated galaxies.     
Errors introduced by projection effects will be discussed later.

\begin{figure*}
 \centering
\includegraphics[width=14cm]{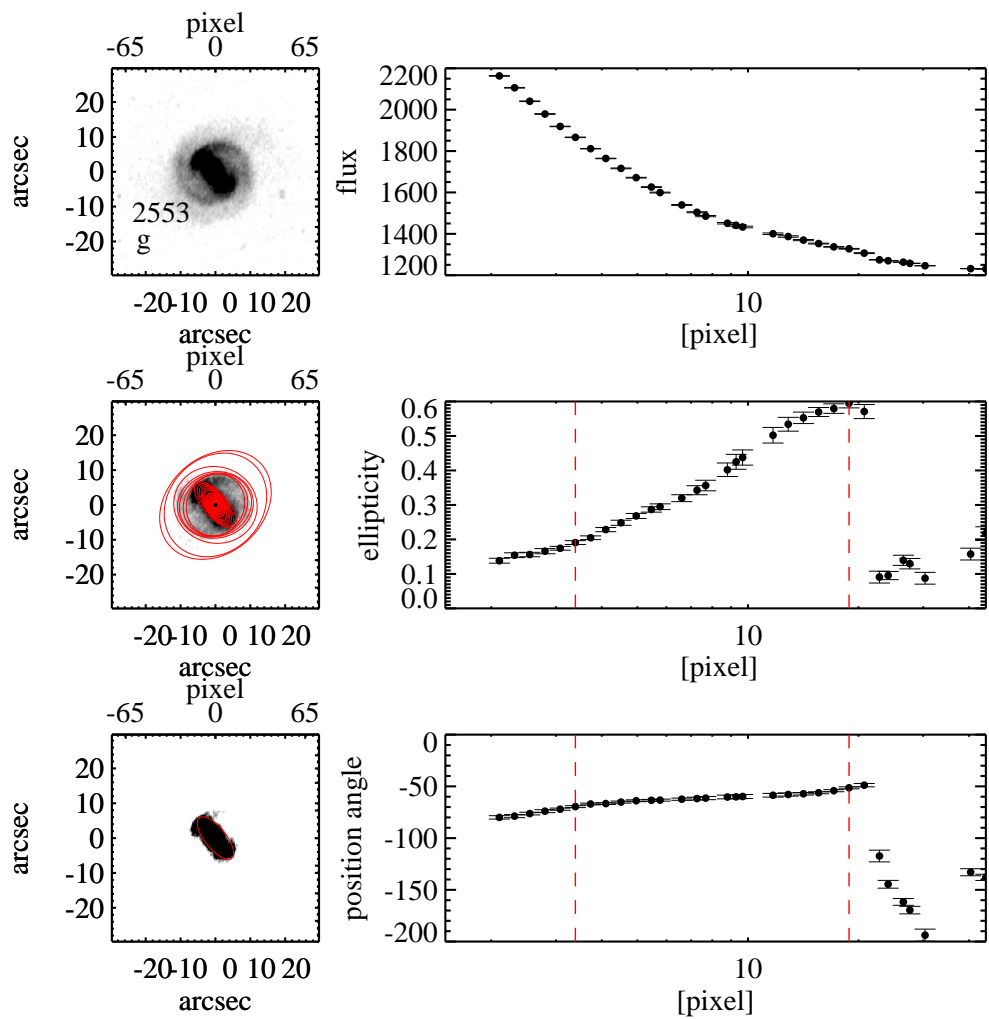}
\caption{An example of a galaxy identified as barred. In the left column,
the top panel shows the image of the galaxy, the middle panel shows the 
fitted ellipses superimposed on the galaxy image, while the bottom panel
shows the same image with the barred region highlighted. 
>From top to bottom, the right column shows the surface brightness, 
ellipticity and position angle profiles of the galaxy.}
 \label{fig:barexample}
\end{figure*}

The criteria we use to classify a galaxy as ``barred'' are similar to those adopted
by  \citet{Jogee04}, hereafter J04, with a few changes
that are detailed below.  J04 required that (1) within the bar, the ellipticity
should rise to a  maximum value greater than 0.25,
 and that  the position angle within the bar should remain constant to within  20
degrees; 
(2) at the end of the bar, the ellipticity should  drop by more than 0.1 and the
position angle 
should change by more than 10 degrees.  Figure~\ref{fig:barexample} illustrates  the
ellipticity
and position angle profiles of a galaxy that is classified as barred according to
these criteria.
 
We  imposed the following additional requirements after visually inspecting
all the fits : (1) the inner radius of the bar should be less than 0.5 times the outer
radius of the bar, and  ellipticity should increase monotonically as a function of
radius from  0.33 $R_{outer}$ to the  end of the bar. This requirement improves the
ability of the algorithm to
identify bars in galaxies with prominent bulges 
\citet{Martin95} showed that  the  size of the bulge
is on average about   half the size of the bar). 
It also prevents complicated spiral
structures and inner rings
from being mistakenly identified as bars.
(2) The outer radius of the bar should be less than 0.8 times the semi-major axis of
the outermost fitted ellipse. 
This requirement ensures that edge-on, disk-dominated galaxies, where the
inclination has been
incorrectly measured, do not enter our sample of barred galaxies.   
(3) if  criterion 1 adopted by J04  is satisfied, and the 
ellipticity drops by more than 0.25 at the end
of the bar,  but the position angle does not change by more than  10 degrees, we
still classify the galaxy as  barred. 

The final requirement ensures that we do not exclude galaxies which have a long bar
surrounded by an aligned ring 
Examples of such systems are given in  \citet{Gadotti03}. 
We also note that \citet{Menendez07} estimated that 
the J04 position angle criterion could introduce incompleteness
at the  10\% level.  
Criterion 3  
resolves this problem for  bars which have high ellipticity ($e_{bar}>$0.5),
which are the focus of our study.

As well as classifying galaxies into barred and non-barred systems, we compute two
quantitative parameters: 1) the bar ellipticity $e_{bar}$, defined as the 
value of the ellipticity at the end of the bar, 
and 2) the relative bar size $D_{bar}/D_{disk}$, where $D_{bar}$ is the diameter of
the bar and $D_{disk}$ is
is the diameter of the 25 mag$/arcsec^2$ isophote measured in the $g$-band. 
The median value of $e_{bar}$ is 0.47 and the median value of $D_{bar}/D_{disk}$ is
0.3 for the galaxies in this study.

The most distant galaxies in our sample are at $z \sim 0.05$, where 2.5 times the
FWHM of the SDSS PSF ($\sim3.5$ arcsec) corresponds to
a physical scale of  3.8 kpc.   Note that a 3.8 kpc bar aligned along the minor axis 
of an inclined galaxy with  $1-b/a=$0.25 has a de-projected
physical size of 5 kpc, so our sample is complete for bars 
larger than 5 kpc. 

\begin{figure*}
 \centering
\includegraphics[width=7cm]{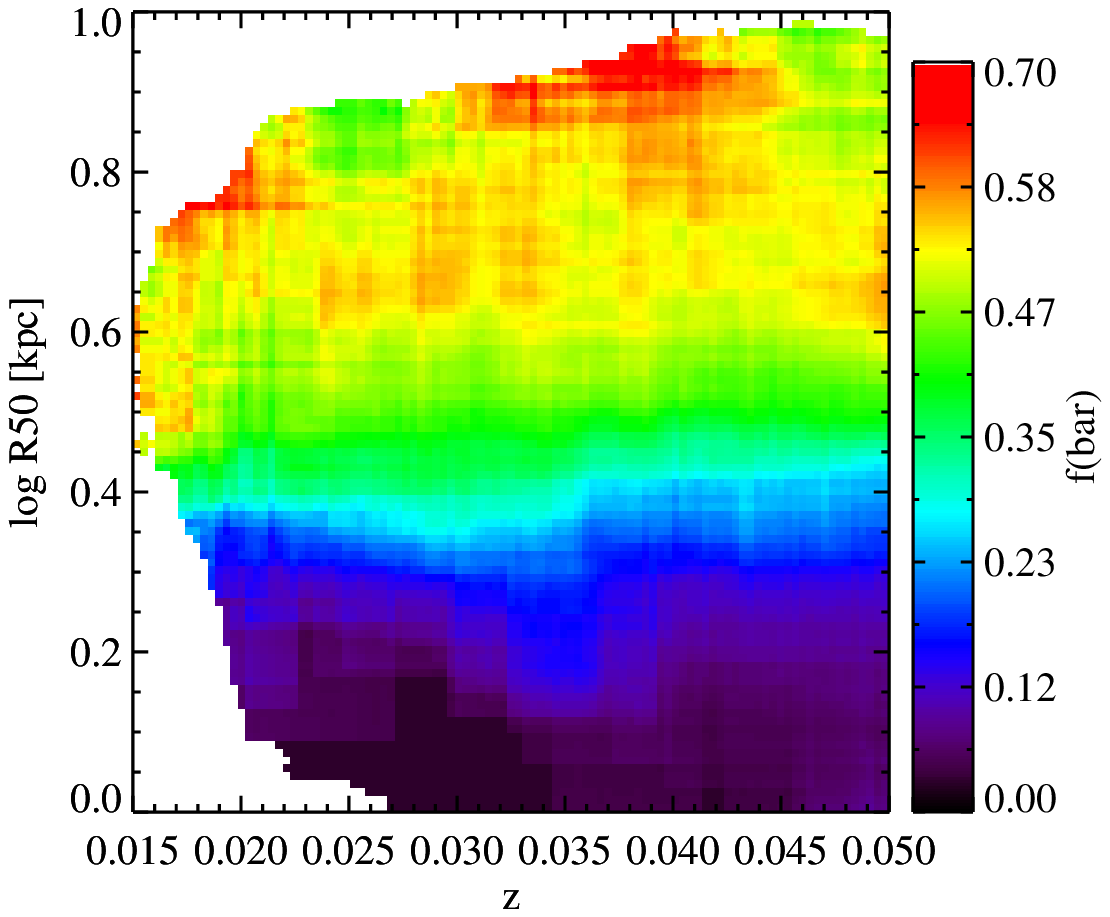}
\includegraphics[width=7cm]{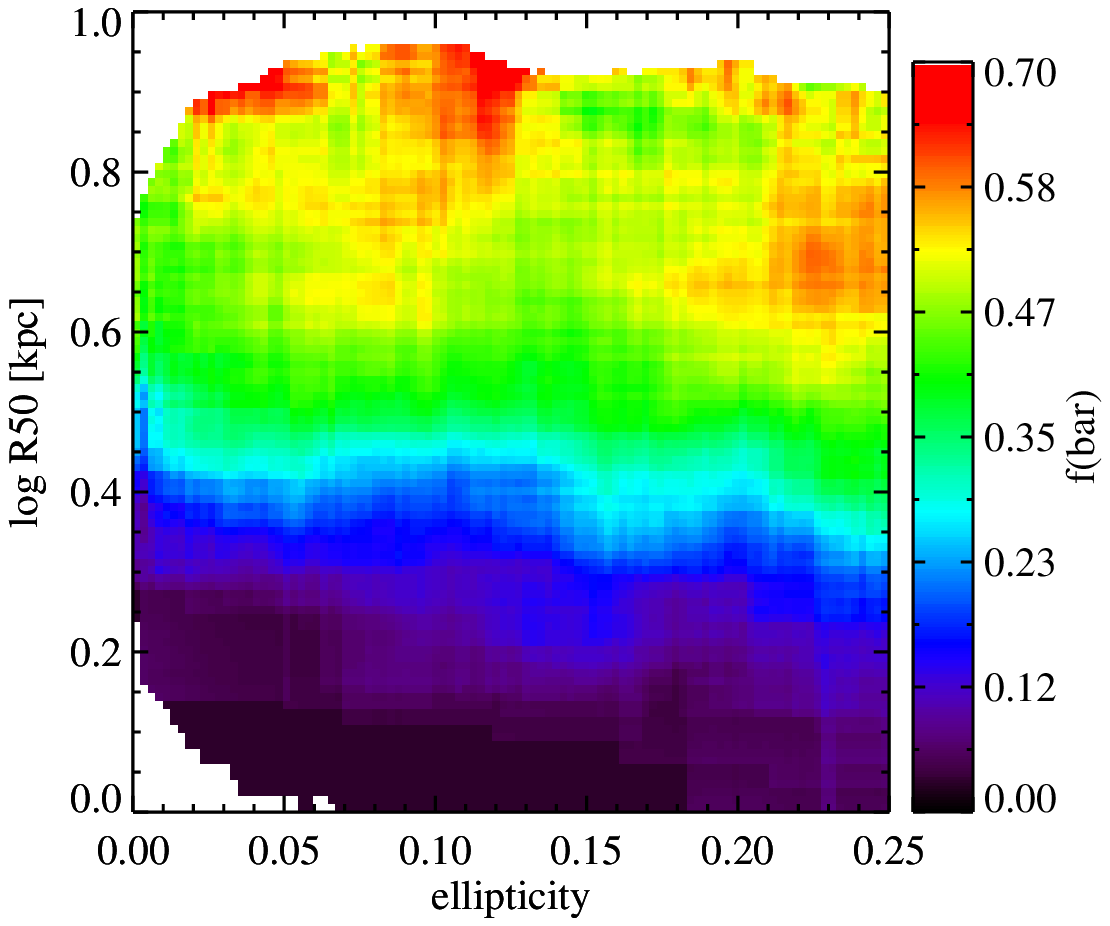}
\caption{The distribution of galaxies in our sample is plotted in the plane 
of physical size versus redshift (left), and physical size versus ellipticity
(right). The coloured contours denote the fraction of bars as a function
of position in this plane. Only Sample A barred galaxies are plotted.} 
 \label{fig:barcheck}
\end{figure*}

To demonstrate that our 5 kpc cut produces an unbiased sample of barred galaxies, we
plot  the fraction of barred systems
as a function of redshift and galaxy size in 
the left panel of Figure~\ref{fig:barcheck}. If the barred galaxy sample suffers from
incompleteness because we miss bars with smaller angular sizes, we
might expect the bar fraction to drop at higher redshifts, particularly
in small galaxies. We see, however,  that bar fractions do not vary  with redshift
at fixed physical size. The right panel of Figure~\ref{fig:barcheck} shows  the
fraction of 
barred galaxies as a function of galaxy size
and ellipticity.  As can be seen, the bar fraction does not vary 
with ellipticity, so 
there is no clear inclination-dependent bias in our estimates
of bar fraction, at least for values of $e_{disk}$ less than 0.25.
We will refer to the sample of galaxies with bars larger than 5 kpc as 
Bar Sample A.

One might worry, however, that by adopting a fixed cut in the physical diameter
of the bar, one would be able to identify a larger proportion of the bars
present in large galaxies compared to smaller ones. Another possibility is
to adopt a fixed cut on $D_{bar}/D_{disk}$. We have examined the distribution
of the diameters $D_{25}$  of the galaxies in our sample. We find a lower
limit of $D_{25}$ of around 16 kpc, corresponding to the diameter of the
most compact galaxies with stellar masses of $10^{10} M_{\odot}$.
We have therefore constructed a second sample of barred galaxies   
with $D_{bar}/D_{disk}>0.3$, which we will refer to as Bar Sample B.

In total, we identify  1555 barred galaxies with
$D_{bar}> 5$ kpc ( $41.4\%$ of the original sample of 3757
galaxies), of which 827 ($22\%$)  have $e_{bar}>0.5$. We note that 
a similar fraction was found by \citet{Marinova07} and \citet{Jogee04} when  
a similar cut in bar size and bar ellipticity was applied.
Bar Sample B contains 506 galaxies, and 363 of them have $e_{bar}>0.5$.
In general, Bar Sample B is most robust for  investigating how
the fraction of bars depends on galaxy structural parameters, such as
stellar surface mass density or concentration. Bar Sample A is most useful for 
investigating the fraction of central starbursts that were triggered by
bar-induced inflows.

Another issue that may potentially bias our results is that if the galaxy has a bulge,
it may be more difficult to measure the shape of
the bar by means of the ellipse-fitting technique. 
The bulge will make the surface brightness contours
of the galaxy rounder, causing $e_{bar}$ to be under-estimated. As a result,
some barred galaxies will then be missing from our  sample.  
One way we can quantify this effect is through simulations, which are
described in detail in Appendix A.
The main conclusion of this study  is that the identification of ``strong bars'' with  
$e_{bar}> 0.5$ is reasonably robust to the presence of a bulge. The
interested reader is referred to the Appendix for more details.
In the rest of the paper, our  conclusions will be based on
the analysis of bars with $e_{bar}> 0.5$. 

\subsection{Control samples}

There have been many papers that have found that barred spiral galaxies have higher 
central star formation rates than un-barred 
spirals \citep{deJong84, Devereux87, Hawarden86, Ho97, 
Martin94, Puxley88, Zaritsky94}. The enhancement
is more pronounced 
for  strongly barred galaxies than for weakly barred 
galaxies \citep{Ho97, Martin95, Martin94}. 

It should be noted that the star formation rates in both 
barred and un-barred galaxies 
vary strongly as a function of their structural parameters,
such as concentration and stellar mass surface density.   
If we wish to quantify the enhancement in central star formation rate accurately,
we need to create control samples that are closely matched with  the barred galaxy
sample in these parameters. 

For each barred galaxy from Sample A, 
we search the parent sample for a matching galaxy (selected without regard
to whether or not it has a bar)  
with stellar mass difference $\Delta \log M_*$  
less than  0.15, stellar surface mass density difference  $\Delta \log \mu_*$
less than 0.1,  and $g-i$ colour difference $\Delta (g-i)$ less than 0.1.

\subsection{Photometry of bars and pseudo-rings}

As described in Section 2.2, the outer end of a bar is defined as the point where the
ellipticity profile reaches its maximum. From now on, we will refer to
the region within the ellipse enclosing the end of the bar as the 
``barred region'' of the galaxy. 
According to the simulations of \citet{Ath92}, gas will flow from the end of
the bar towards the center of the galaxy, following a  
curved path which is determined by the strength of the 
bar and the central mass density of the galaxy. 
The region of the galaxy that we define to be barred will cover the path 
of the inflowing gas until it condenses
into a ring at the inner Lindblad resonance (see Figure 3).

We also define an outer ellipse with  ellipticity equal to that
of the disk measured at its  25 $mag$ $arcsec^{-2}$ isophote in the $r$-band and with
major axis equal to  $1.2 D_{bar}$. The ``pseudo-ring'' region of the galaxy
is the defined as the region between the ellipse enclosing the barred region
of the galaxy, and this outer ellipse. 
The motivation for defining such a ring is that an ultra-harmonic 
resonance ring in disk galaxies is often found at the end of the bar
 \citep{Ath09,Ath09b,Ath10,Regan02,Regan03}. 
Gas will tend to accumulate in this ring
and if the density reaches high enough values, stars will form.
These outer resonance rings are usually aligned parallel to the bar and
may extend out to the corotation radius of the disk \citep{Buta10, Schwarz81}. 
The ``pseudo ring'' region defined in this study is thus designed to cover 
the location of  potential resonance rings in our galaxies. 

For each barred galaxy in our sample, we measure the SDSS $g-i$ colour
within the barred region and within the "pseudo ring" region. 
We also measure colours within the same regions in the corresponding control galaxy,
even though these systems often do not have bars see the right panel of figure 3). 
(Note that when we define the corresponding barred and pseudo-ring regions
of the control galaxy, we scale the major and minor axes of both
regions by the ratio  $D_{25}$(control galaxy)$/D_{25}$(barred galaxy),
so that the outer radii of the two systems match exactly)\footnote[1]{We test two ways of measuring colour from
a ``barred region'' (defined by the barred galaxy) in the control
galaxy. In the first way, we do not change the orientation
of a ``barred region'' and in the second way, we orientate
the ``barred region'' so that its major axis points to the same direction
as the major axis of the disk. These two types of colour measurements yield very
similar results. In what follows, we present results using the first set of 
colour measurements.}.   We will compare the colours between the barred galaxies and control galaxies in Section 3.2. 

All the colours in the barred and `pseudo ring'' regions 
are corrected for dust attenuation using the average of the central
attenuation and the global attenuation.
The central attenuation is  derived from the Balmer decrement measured from the
SDSS fibre spectrum.   
We have assumed a \citet{Calzetti00} extinction curve and the  
line-to-continnum attenuation ratio from \citet{Wild11}. 
We apply a dust correction when both
the H$\alpha$ and H$\beta$ fluxes have $S/N>3$. If the emission lines
are too weak to measure the Balmer decrement, we adopt $\tau_V(star, fiber)$ 
from the measurements of the attenuation of the stellar continuum 
of the galaxy provided in the  MPA$\slash$JHU catalog ({\it tauv\_cont}). 
This parameter is obtained by fitting \citet{Bruzual03} population 
synthesis models to the stellar continuum; the reddening may then be 
estimated by determining the extra ``tilt'' that
must be applied to the models in order to fit 
the shape of the observed spectrum. The global attenuation is derived using
SED-fitting techniques.  

\begin{figure*}
 \centering
\includegraphics[width=16cm]{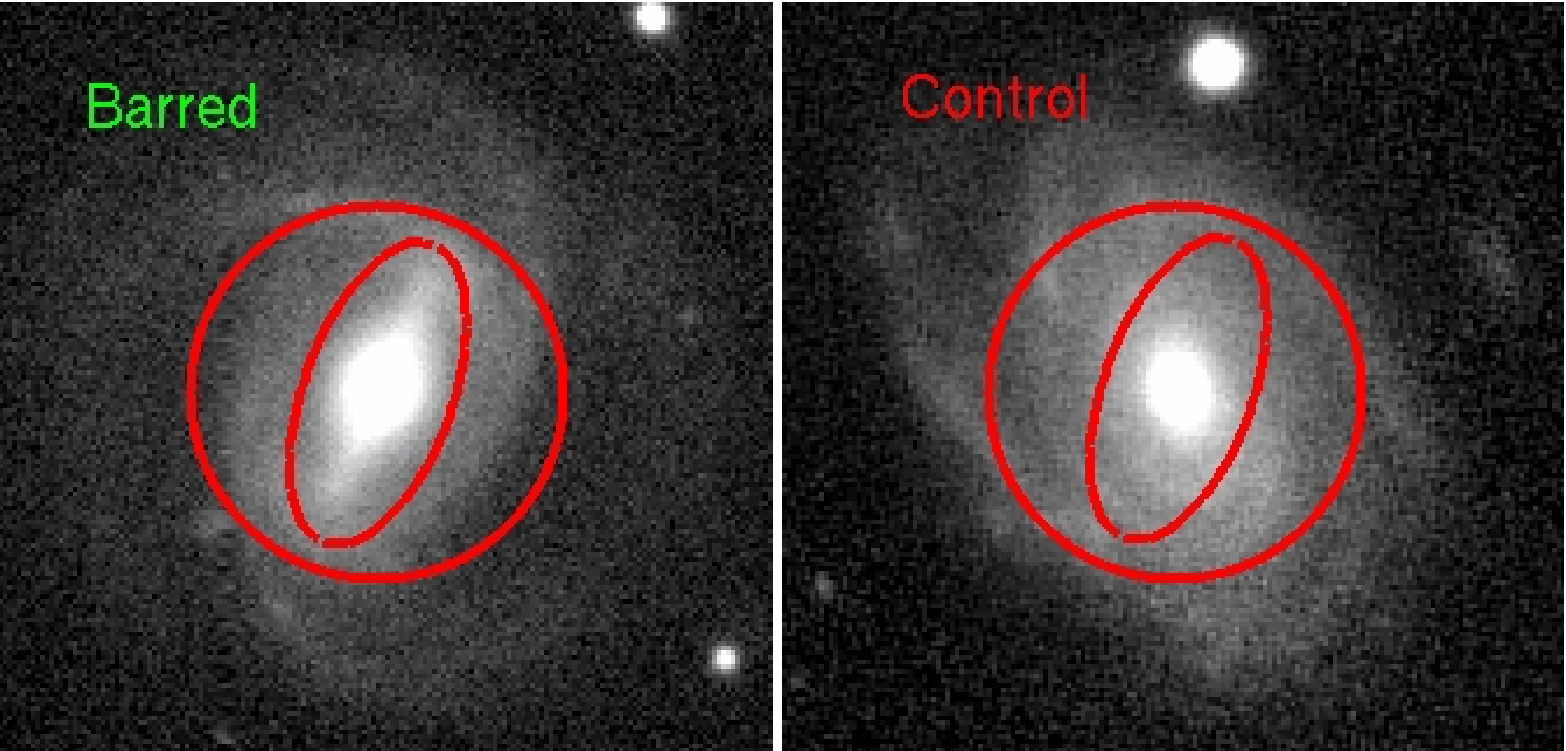}
\caption{An example of a barred galaxy (left) and its matched control galaxy (right). 
In the left panel, the inner ellipse encloses the ``barred'' region of the
galaxy (see text). The area between the inner and outer ellipses is referred to as the
pseudo-ring region. In the right panel, the same inner and 
outer ellipses are plotted on     
the image of the control galaxy.}
 \label{fig:ctrlexample}
\end{figure*}

\section{Results}
\subsection{Establishing a link between strong bars and enhanced central star
formation in galaxies}

As discussed in Section 2.3, we have constructed a control sample
with the same stellar masses, structural properties and global
colours as the barred galaxies. By examining whether the central star formation rates
are enhanced in the barred galaxy sample when  compared to the controls, we can test 
whether there is a causal link between  bars and central starbursts. 
By binning the barred galaxy sample and the control sample according to stellar mass, 
stellar surface density and morphological type, we ascertain to what extent the 
degree of central star formation enhancement is influenced by global galaxy
properties. 

In this paper, we use the quantity sSFR$_{fib}/$sSFR$_{tot}$, which we will
denote C(SF),  as a measure of
how concentrated the star formation is in a galaxy: sSFR$_{fib}$ is the 
specific SFR measured within the 3 arcsec SDSS fiber, and sSFR$_{tot}$ is the 
global specific star formation rate of the galaxy.

\begin{figure*}
\centering
\includegraphics[width=15cm]{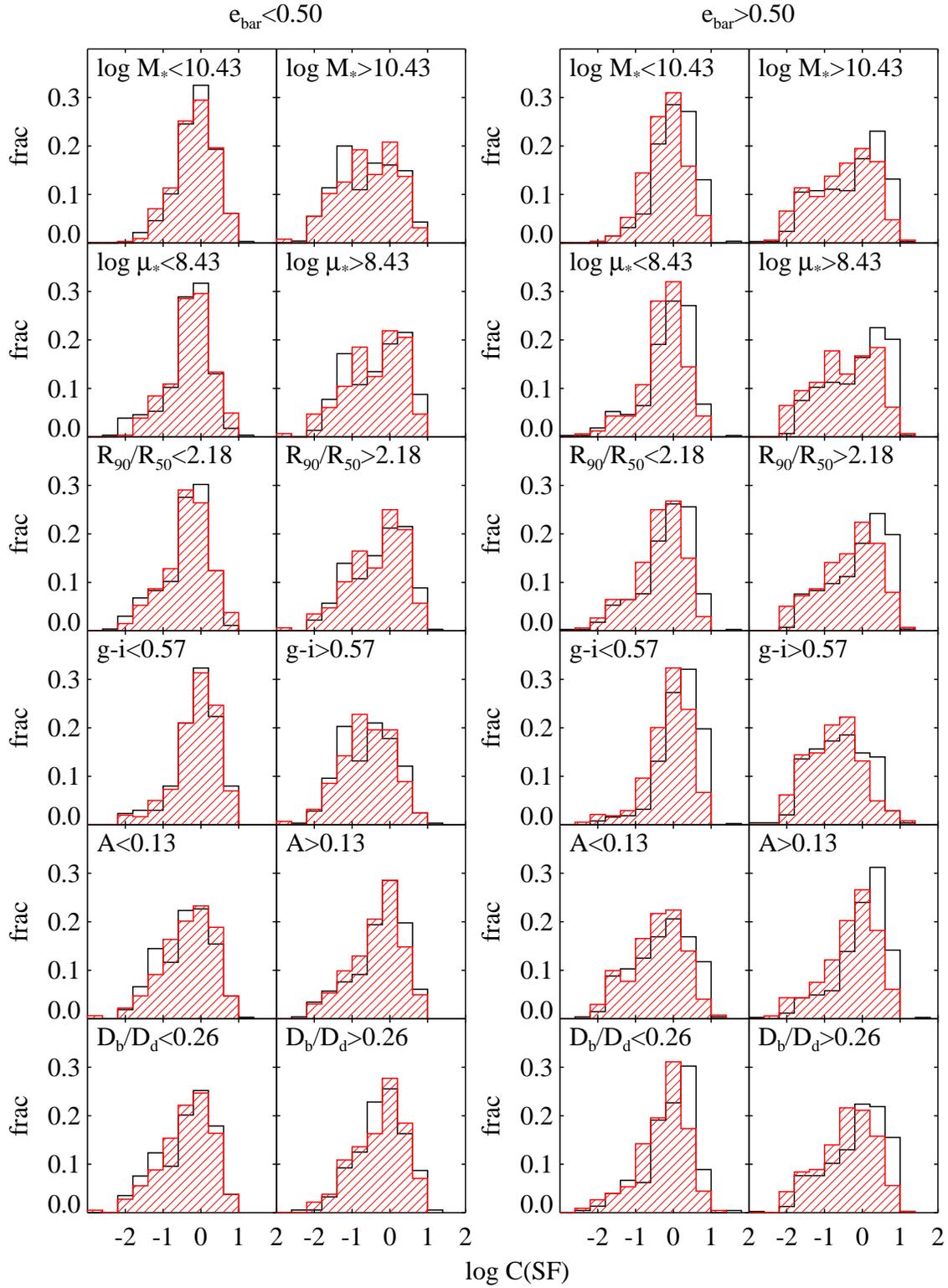}
\caption{ Histograms of the ratio of specific SFR evaluated within the SDSS 
fiber to specific SFR for the galaxy as a whole ($\log$ C(SF))
are shown for the barred galaxy sample (open histograms), and
for the  control sample matched in stellar mass, stellar mass surface 
density and $g-i$ colour (red, hatched histograms). The plots
on the left show results for galaxies with
weak bars ($e_{bar}<0.5$), while
the plots on the right are for galaxies with 
strong bars ($e_{bar}>0.5$). The samples  are further divided 
by stellar mass ($\log M_*$, $M_*$ is in units of $M_{\odot}$),
stellar mass surface density ($\log \mu_*$, $\mu_*$ is in 
units of $M_{\odot}/kpc^2$), concentration ($R_{90}/R_{50}$), 
$g-i$ colour, asymmetry index ($A$) and relative 
bar size ($D_{bar}/D_{disk}$) at the median value of
each parameter,  as denoted in each plot. Only Bar Sample A galaxies are plotted.} 
\label{fig:bar_ssfrcons}
\end{figure*}

In Figure~\ref{fig:bar_ssfrcons}, we compare the distribution of  
$\log$ C(SF) for our barred galaxies (shown as black histograms
on the plot) and control galaxies (red dashed histograms). 
The two columns on the left show results for galaxies with weak bars
($e_{bar} < 0.5$), while the columns on the right are for strong bars
($e_{bar} > 0.5$). Each row in the figure shows results for galaxies split
according to a set of global galaxy properties that include stellar mass,
stellar surface mass density, concentration parameter, $g-i$ colour, 
galaxy asymmetry index $A$, and ratio of the diameter of the bar to
that of the disk ($D_{b}/D_{d}$). In each case, the split is made at the
median value of the parameter under investigation. The difference in the average $\log$ C(SF) and Kolmogorov-Smirnov probability for the barred 
and control galaxies for each sub-sample in each plot are sumarized in Table~\ref{T:CSF_ctrl}.

Our conclusion is very simple. For each of the parameters shown in the two left-hand panels ($e_{bar}<0.5$), there is no significant
difference between the black and red histograms, i.e., 
there are no significant differences between the 
barred and the control samples. For each of the parameters shown in the two right-hand panels ($e_{bar}>0.5$),
the differences ara very significant. This shows that only strong bars are 
capable of inducing enhanced central star formation in galaxies. 
In the right two panels, the offset between the black and red histograms 
is roughly the same for each parameter. This shows that the global galaxy properties considered here
do not play a significant role in determining the degree of central
star formation enhancement caused by bar-driven inflows.

\begin{figure*}
 \centering
\includegraphics[width=5cm]{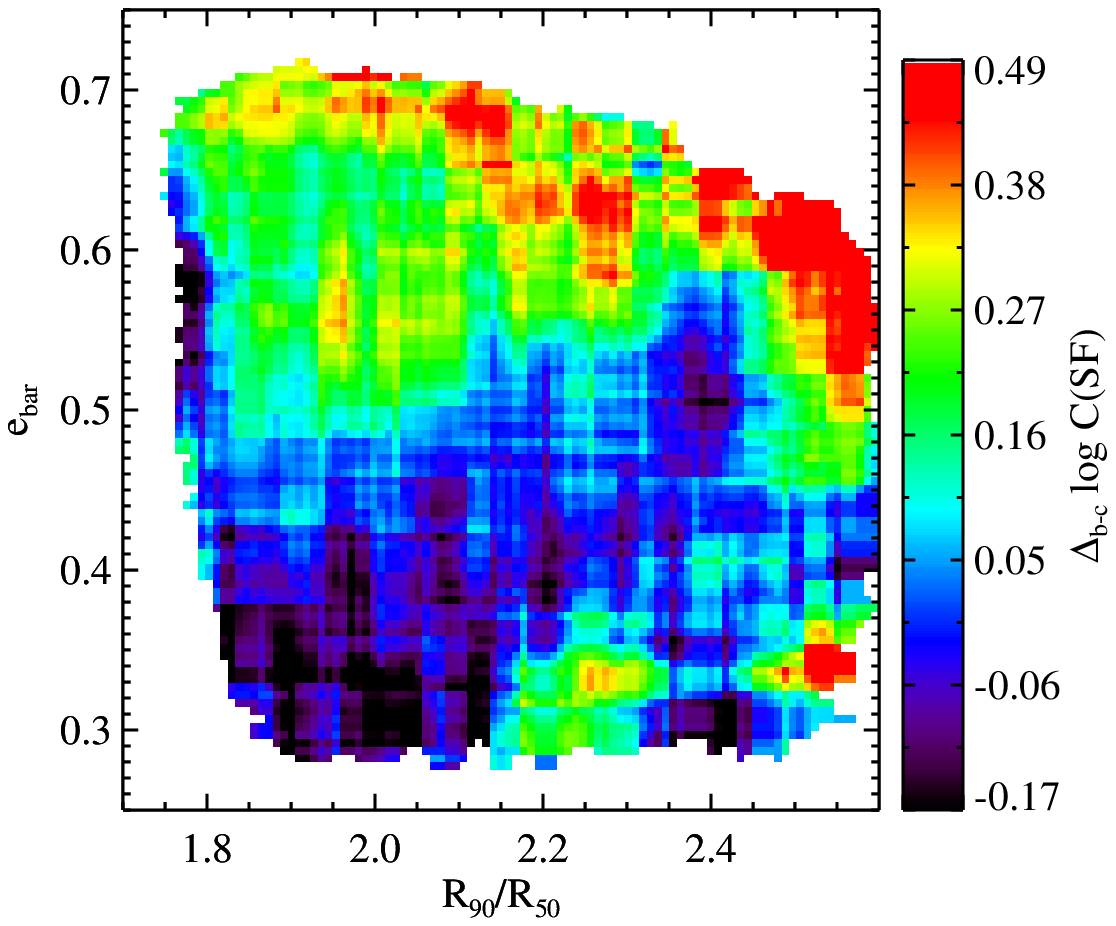}
\includegraphics[width=5cm]{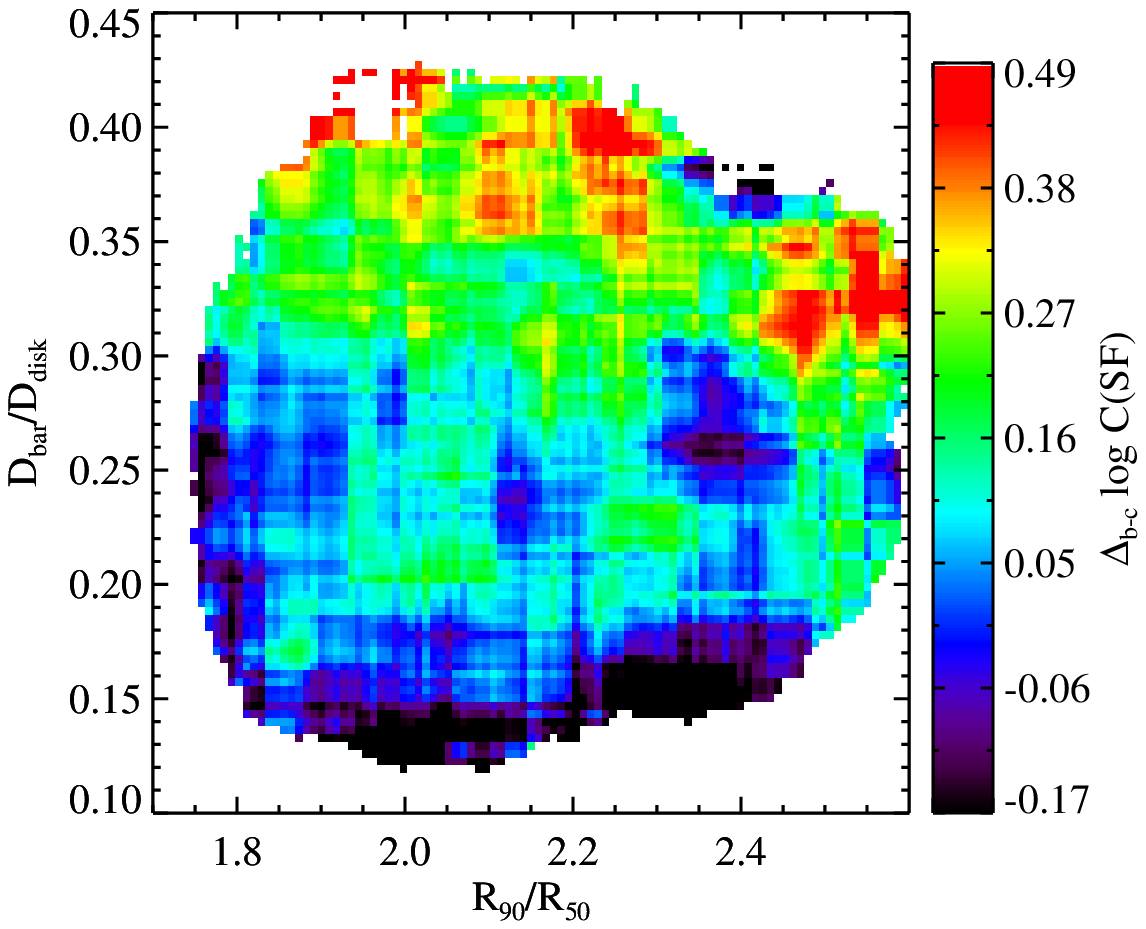}
\includegraphics[width=5cm]{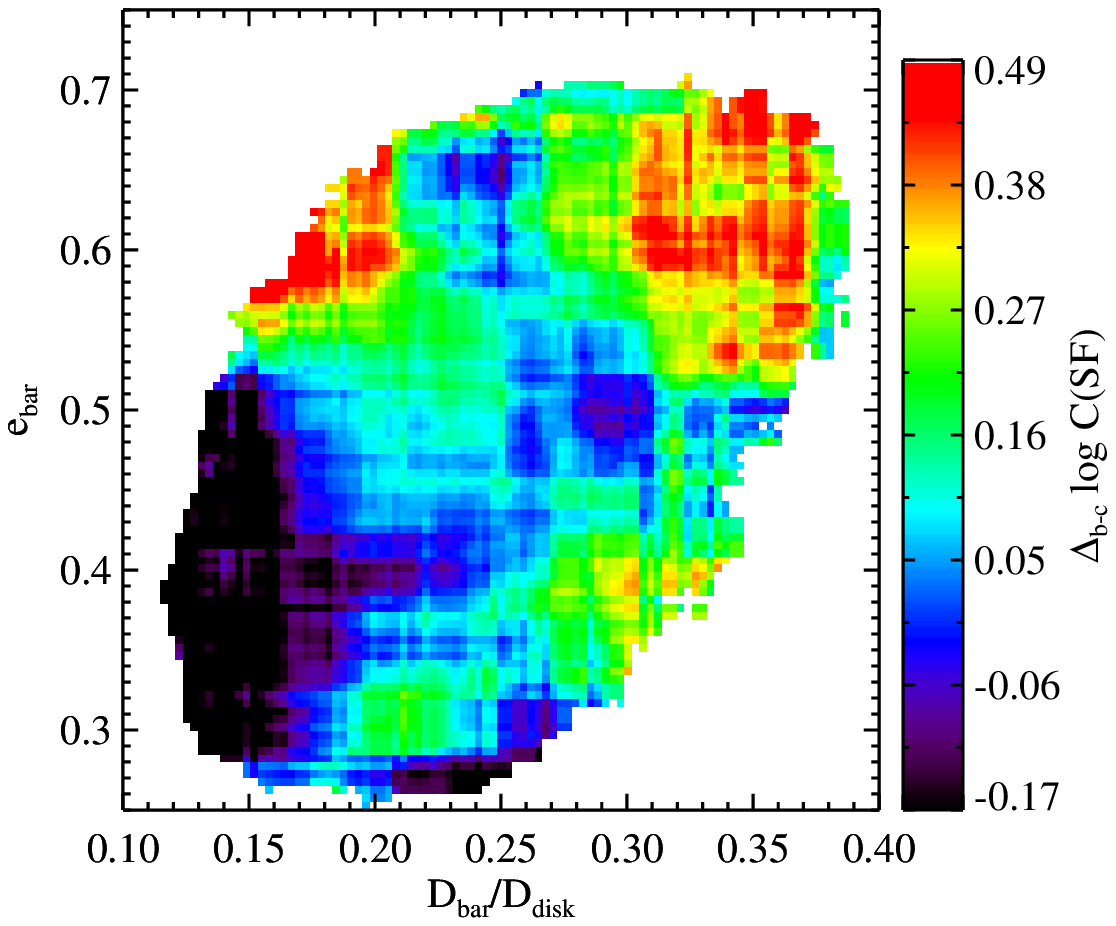}
\caption {The distribution of C(SF) of barred galaxies minus C(SF) of control sample
galaxies is plotted in the plane of bar ellipticity parameter
$e_{bar}$ versus concentration (left), bar relative size parameter
$D_{bar}/D_{disk}$ versus concentration (middle) and $e_{bar}$ versus bar relative size parameter (right).
Coloured contours indicate
the average value of  C(SF, barred)-C(SF, control)
as a function of position in this plane.  Note that the control galaxies 
are assigned the values of $e_{bar}$ and $D_{bar}$ of 
their corresponding barred galaxies when making the plots. Only Bar Sample A galaxies are considered.}
 \label{fig:D_ssfrcons_cons}
\end{figure*}

In the left panel of Figure~\ref{fig:D_ssfrcons_cons}, we show  how the degree of central
star formation enhancement depends on both  $e_{bar}$ and on the concentration
index of the galaxy (note that the concentration index is a simple measure
of  bulge-to-disk ratio).  
We plot contours of $ \Delta_{b-c} \log $C(SF) in the plane of
$e_{bar}$ versus R$_{90}$/R$_{50}$, where $\Delta_{b-c} \log$ C(SF) is the
difference in  $\log $[sSFR$_{fib}/$sSFR$_{tot}$] for the barred galaxies
and the controls.   
We see that when $e_{bar}>0.5$,  $ \Delta_{b-c} \log $C(SF) is positive,
indicating that star formation is more concentrated in the barred galaxies.  
When $e_{bar}<0.5$, star formation in barred galaxies and control galaxies
are similarly concentrated.

We note that there is an apparent deficit of galaxies with large concentrations
and with high values of $e_{bar}$ in Figure~\ref{fig:D_ssfrcons_cons}. 
As discussed in Appendix A,  one possible reason for this 
is because $e_{bar}$ will be systematically underestimated in galaxies
with larger bulge-to-disk-ratios. In Figure A2, we see that $e_{bar}$
is underestimated by about 0.1-0.15 when $R_{90}/R_{50} > 2.5$, which agrees
well with the observed deficit.  
After taking this bias into account, we conclude that 
there is no clear evidence that strong bars are less efficient at 
channeling gas to the central regions of galaxies if a  
bulge component is already present.  On the other hand,
Figure~\ref{fig:D_ssfrcons_cons} show that the degree of central
star formation enhancement appears to be {\em smaller}
in strongly barred galaxies with low central mass concentration ($C< 1.9$),
in agreement with simulation results from  \citep{Ath92} and
\citet{Sheth00}. 

In the the middle panel of Figure~\ref{fig:D_ssfrcons_cons}, we investigate whether
the degree of central star formation enhancement exhibits any dependence on
the length of the bar relative to that of the disk. We plot 
 $\Delta_{b-c} \log$ C(SF) in the plane of   
$D_{bar}/D_{disk}$ versus concentration index. We see that  
$ \Delta_{b-c} \log$ C(SF) is somewhat  higher when $D_{bar}/D_{disk}>0.3$. 
We note however, that this effect is mainly due to the fact that 
the size of the bar is correlated with its ellipticity.            
This is clearly demonstrated   
in the right panel of the figure, which shows contours of $\Delta_{b-c} \log$ C(SF) in 
the $D_{bar}/D_{disk}$ versus $e_{bar}$ plane. The main dependence of
C(SF) is on $e_{bar}$ and not on $D_{bar}/D_{disk}$. 

In summary, we find that the central star formation enhancement
depends mainly on the ellipticity of the bar, and not  on 
the size of the bar or on the  mass or  structure of the host galaxy.

\begin{table*}
\begin{center}
\begin{tabular}{|c|cc|cc||cc|cc|}
\hline
 \multicolumn{1}{|c|}{} & \multicolumn{4}{|c|}{$e_{bar}<0.5$} & \multicolumn{4}{||c|}{$e_{bar}>0.5$} \\

\hline

\multicolumn{1}{|c|}{X} & \multicolumn{2}{|c|}{$X<median(X)$} & \multicolumn{2}{|c|}{$X>median(X)$} & \multicolumn{2}{||c|}{$X<median(X)$} &    \multicolumn{2}{|c|}{$X>median(X)$} \\

\hline
 \multicolumn{1}{|c|}{}
&  \multicolumn{1}{|c|}{$\Delta_{b-c}\log C(SF)$} & \multicolumn{1}{|c|}{KS prob} &  \multicolumn{1}{|c|}{$\Delta_{b-c}\log C(SF)$} & \multicolumn{1}{|c|}{KS prob} & \multicolumn{1}{||c|}{$\Delta_{b-c}\log C(SF)$} & \multicolumn{1}{|c|}{KS prob} & \multicolumn{1}{|c|}{$\Delta_{b-c}\log C(SF)$} & \multicolumn{1}{|c|}{KS prob}\\
\hline

$M_*$    & 0.03& 0.14& -0.01& 0.34& 0.23& 0.00& 0.19& 0.00  \\
$\mu_*$  &-0.04& 0.75&  0.07& 0.32& 0.11& 0.00& 0.32& 0.00 \\
$R_{90}/R_{50}$&-0.03& 0.84&  0.05& 0.41& 0.19& 0.00& 0.23& 0.00   \\
$g-i$    &-0.00& 0.93&  0.03& 0.46& 0.26& 0.00& 0.12& 0.09   \\
$A$      &-0.03& 0.47&  0.08& 0.06& 0.11& 0.02& 0.29& 0.00    \\
$D_{bar}/D_{disk}$   
         &-0.01& 0.40&  0.06& 0.82& 0.17& 0.00& 0.23& 0.00    \\
\hline \hline
\end{tabular}
\caption{The difference of the C(SF) between the barred and control galaxies (see Figure~\ref{fig:bar_ssfrcons}). $\Delta_{b-c}\log C(SF)$ is the average value of $\log$C(SF,barred)--$\log$C(SF,control). KS prob is the Kolmogorov-Smirnov probability that the two distributions (C(SF) for the barred and control galaxies) are drawn from an identical parent population. The columns
on the left show results for galaxies with
weak bars ($e_{bar}<0.5$), while
the columns on the right are for galaxies with 
strong bars ($e_{bar}>0.5$). The samples  are further divided 
by stellar mass ($\log M_*$, $M_*$ is in units of $M_{\odot}$),
stellar mass surface density ($\log \mu_*$, $\mu_*$ is in 
units of $M_{\odot}/kpc^2$), concentration ($R_{90}/R_{50}$), 
$g-i$ colour, asymmetry index ($A$) and relative 
bar size ($D_{bar}/D_{disk}$) at the median value of
each parameter,  as denoted in the second row of the table. Only Bar Sample A galaxies are calculated.} \label{T:CSF_ctrl}
\end{center}
\end{table*}

\subsection{Bar and pseudo-ring colours}
\begin{figure*}
 \centering
\includegraphics[width=16cm]{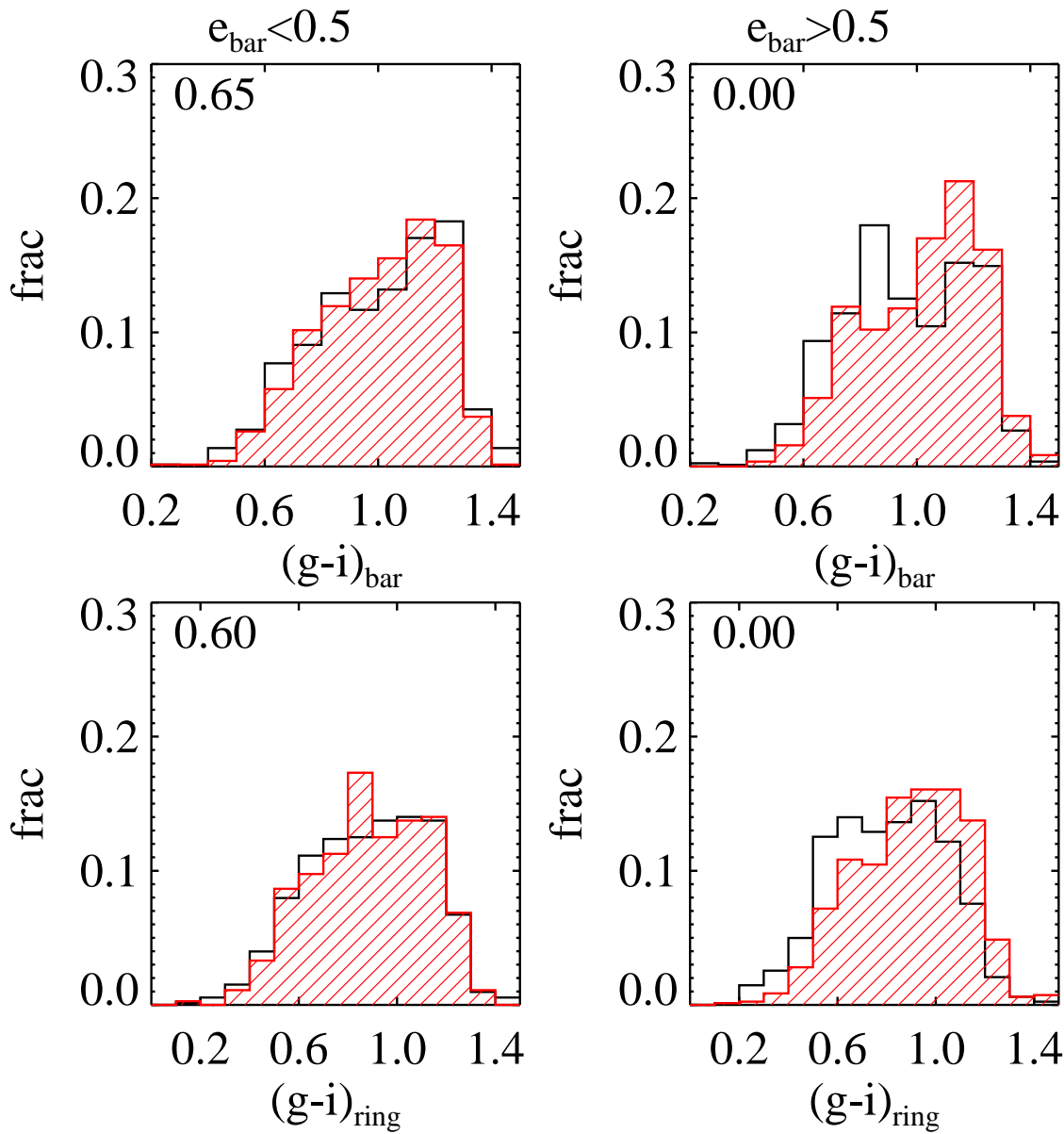}
\caption{The top panels show a comparison of the distribution 
of $g-i$ colours measured for the
`` barred regions''
of the galaxies classified as barred  (black histograms) with the distribution of
$g-i$ colours measured for the same regions in the sample of control
galaxies (red dashed histograms). Results are shown for weak
bars (left), and for strong bars (right). The bottom panels show a comparison
of $g-i$ colours measured for the ``pseudo-ring'' regions of the barred
and control galaxy samples.  Only Bar Sample A galaxies are plotted.}
\label{fig:bar_nonbar_dcl_barstr}
\end{figure*}

In this section, we  compare the colours of the barred and pseudo ring regions 
in the barred and control galaxy samples. 
As shown in the previous section, the degree of central star formation depends  
prinarily on ellipticity, so we simply split the sample 
at $e_{bar}=0.5$.  The left panels of Figure~\ref{fig:bar_nonbar_dcl_barstr}
show that the colours  in the barred and pseudo-ring regions
do not differ significantly from those
measured in the control samples when $e_{bar}< 0.5$. Much stronger
differences are found when $e_{bar} > 0.5$.
These results imply that the enhanced central star formation
extends well beyond the central 
region of the galaxy when the bar is strong. Enhanced  star
formation is found in  regions where gas 
is likely to flow inwards, as well as
in the outer regions of the galaxy, where gas is compressed by the resonance.

\begin{figure*}
 \centering
\includegraphics[width=7cm]{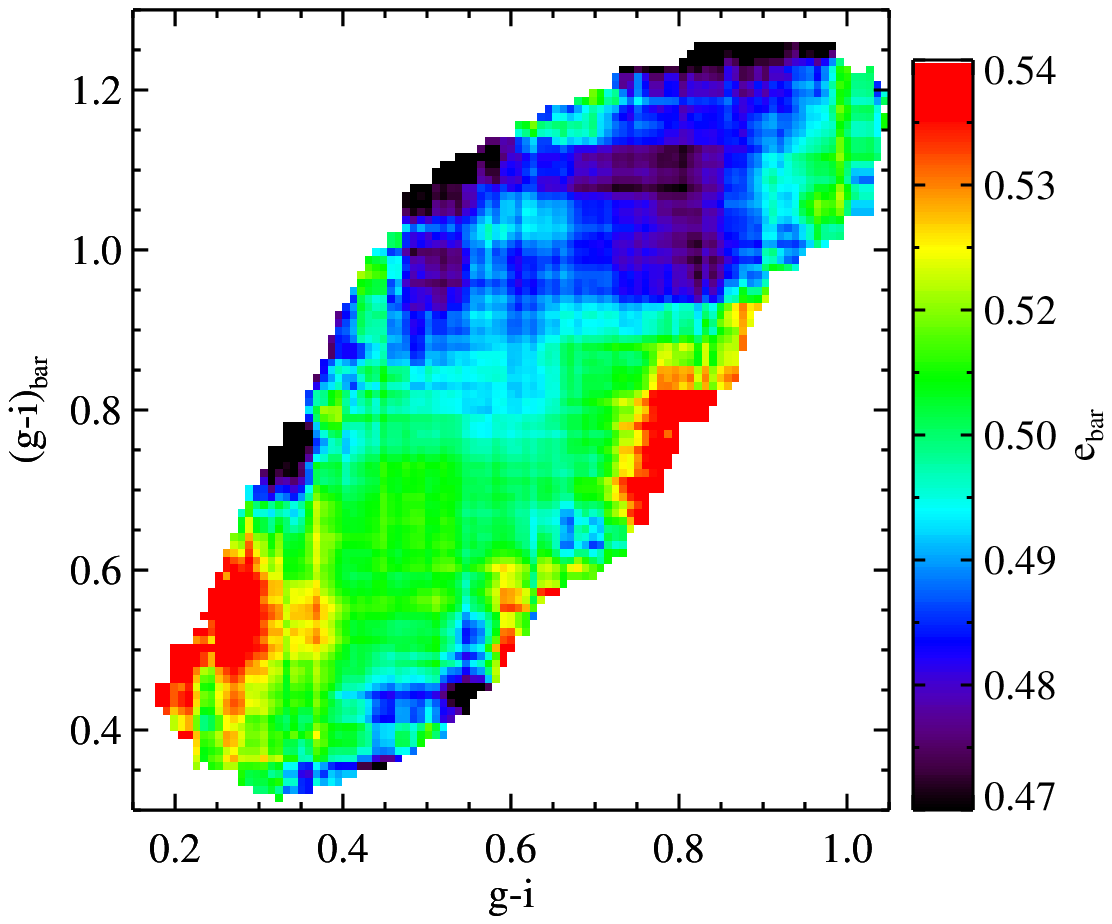}
\includegraphics[width=7cm]{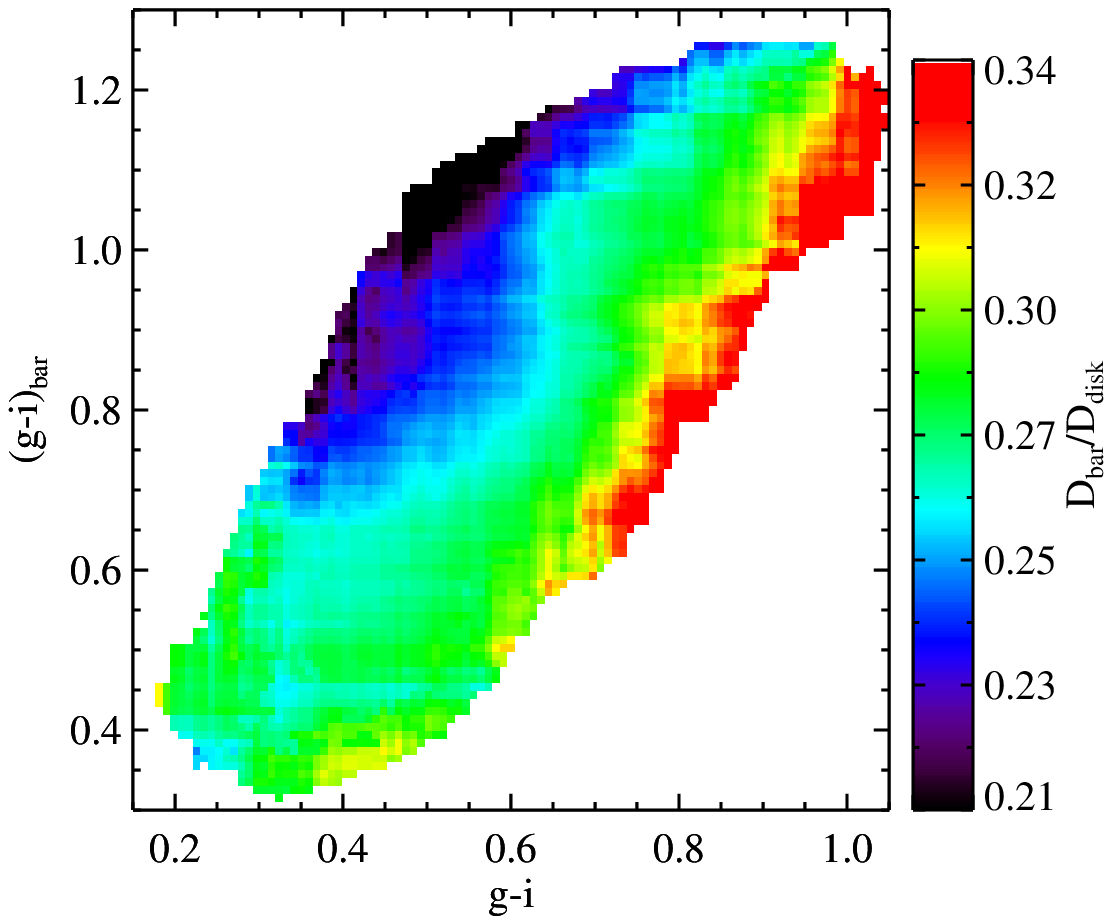}
\caption{The barred galaxy sample is plotted in the 2D plane of bar colour
versus global galaxy colours. In the left panel, coloured contours denote the 
average $e_{bar}$ as a function of position in the plane. In the right panel,
the contours indicate the average value of $D_{bar}/D_{disk}$.  Only Bar Sample A galaxies are plotted.}
 \label{fig:youngbar}
\end{figure*}

Finally, in the left panel of  Figure~\ref{fig:youngbar}, we examine 
how the average ellipticity of the bar varies as a function of location in the
plane of bar colour versus global galaxy colour. As can be seen, if the barred
region of the galaxy is blue, then a strong bar is generally located within it
(note that the ``noise'' in the plot probably arises because the barred region
only  traces the gas inflow region in an approximate way).  
On the other hand, there is no obvious effect seen as a function of the
{\em global} colour of the galaxy.
In the right panel, we examine how $D_{bar}/D_{disk}$ varies as a function of
the same two parameters. In contrast to the results obtained for $e_{bar}$,
the size of the bar (scaled to that of the disk) does not correlate
with the colour of the bar, but scales strongly as
a function of the global colour of the galaxy. The reddest galaxies host the
longest bars. We will discuss possible explanations for this result in Section 4.

\subsection{Fraction of galaxies with strong bars}

In this section, we examine how the fraction of disk 
galaxies with strong bars (i.e. galaxies having $e_{bar}>0.5$)
varies as a function of  stellar mass, stellar mass surface density, 
and concentration.  We show results for 
bar samples A and B. Sample A, which includes all galaxies
with bars larger than a physical size of 5 kpc, 
is useful for deriving a stronger lower limit to the
fraction of galaxies with strong bars. Sample B, which
include only those galaxies with $D_{bar}/D_{disk}>0.3$,  
should yield unbiased trends in  strong bar fraction 
as a function of parameters such
as stellar surface mass density, which scale with the size of the galaxy. 

\begin{figure*}
 \centering
\includegraphics[width=7cm]{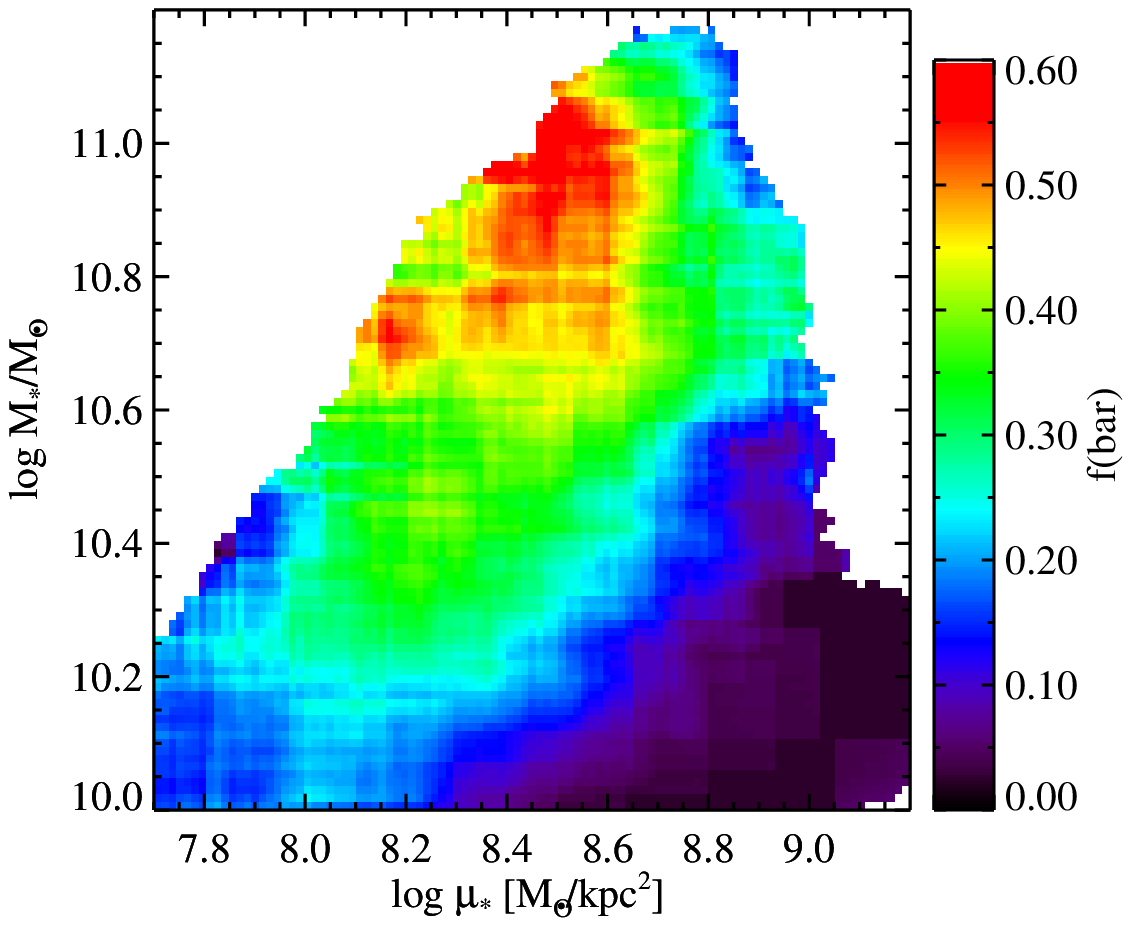}
\includegraphics[width=7cm]{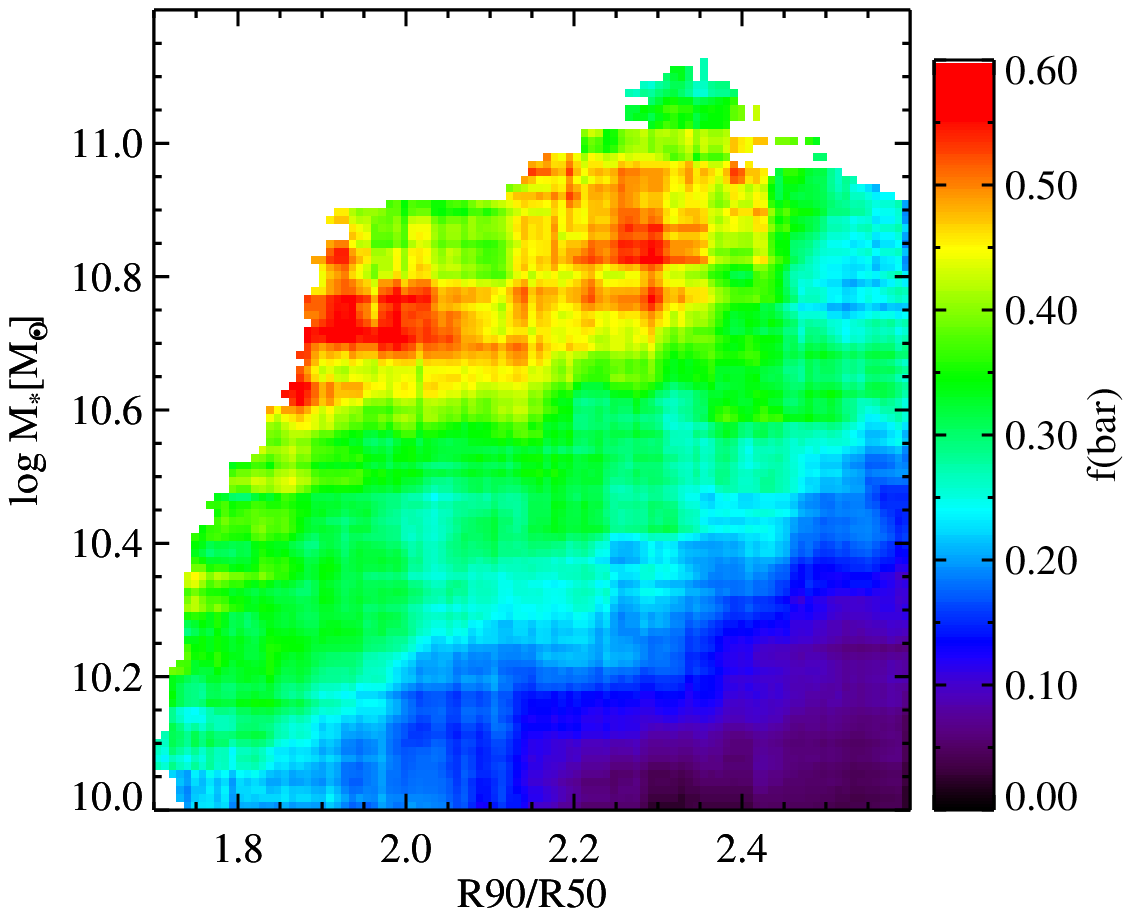}
\includegraphics[width=7cm]{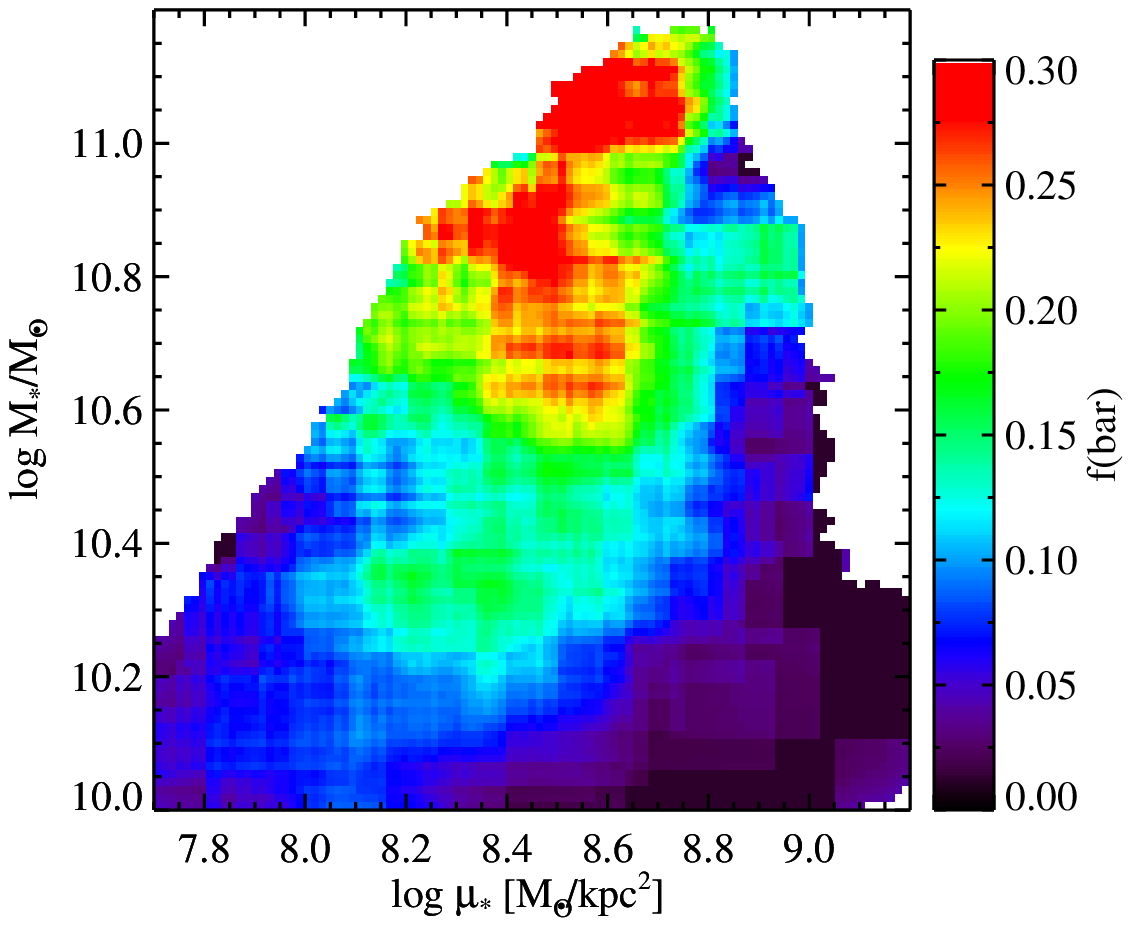}
\includegraphics[width=7cm]{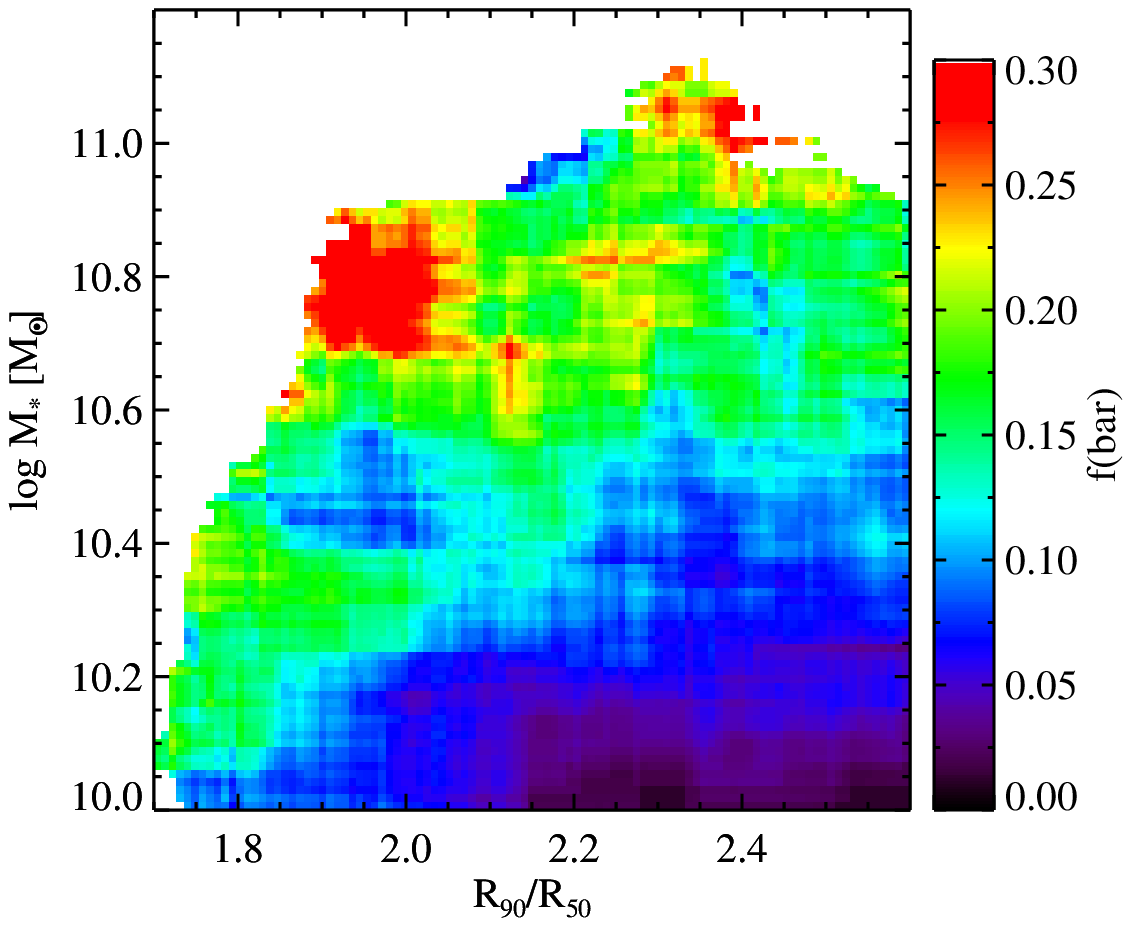}
\caption{The distribution of galaxies from Bar Sample A (top row) and 
Bar Sample B (bottom row) is plotted in the 2-dimensional
plane of  stellar mass versus mass surface density and stellar mass versus concentration. The coloured contours denote the strong bar fraction as a function of position in the plane.}
 \label{fig:strbarfrac}
\end{figure*}

In the top row of Figure~\ref{fig:strbarfrac}, we plot contours of 
strong bar fraction from Bar Sample A in the plane of stellar mass versus stellar surface mass
density (left), and stellar mass versus concentration index (right). 
The fraction of galaxies with strong bars exceeds 0.5  
for galaxies with $\log M_*/M_{\odot}> 10.6$ with  low stellar mass
surface densities ($\log \mu_*/M_{\odot} kpc^{-2} < 8.5$) and low concentrations ($C< 2.5$). 
For lower mass galaxies, the strong bar fraction is a factor of two smaller,
but also peaks at low stellar surface densities and concentrations.

In the bottom row of Figure~\ref{fig:strbarfrac}, we show results for sample B.
The strong bar fractions are lower and the plots are noisier, but the same qualitative
trends are apparent.
We conclude, therefore, that the bar-driven mode of bulge 
formation appears to be most ubiquitous
in massive, disk-dominated galaxies at the present day.

\begin{figure*}
 \centering
\includegraphics[width=5cm]{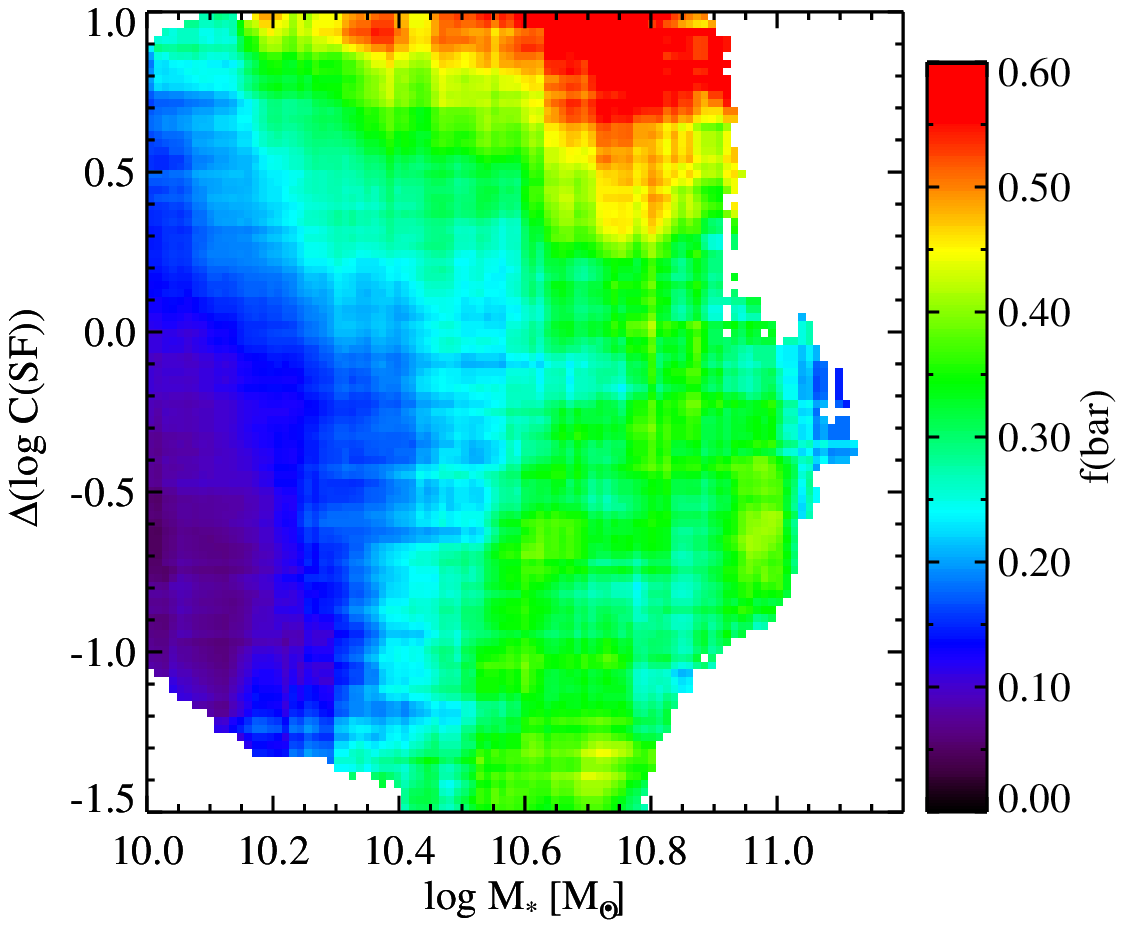}
\includegraphics[width=5cm]{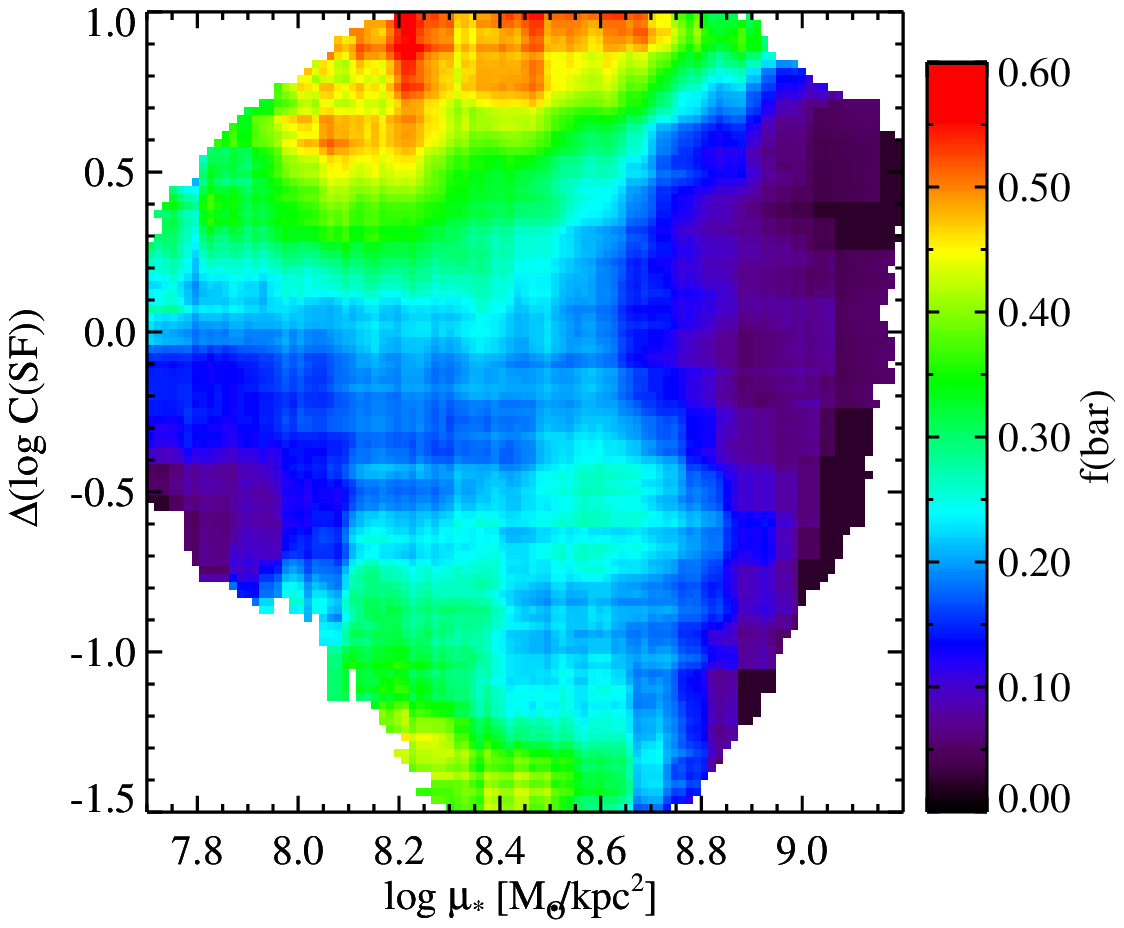}
\includegraphics[width=5cm]{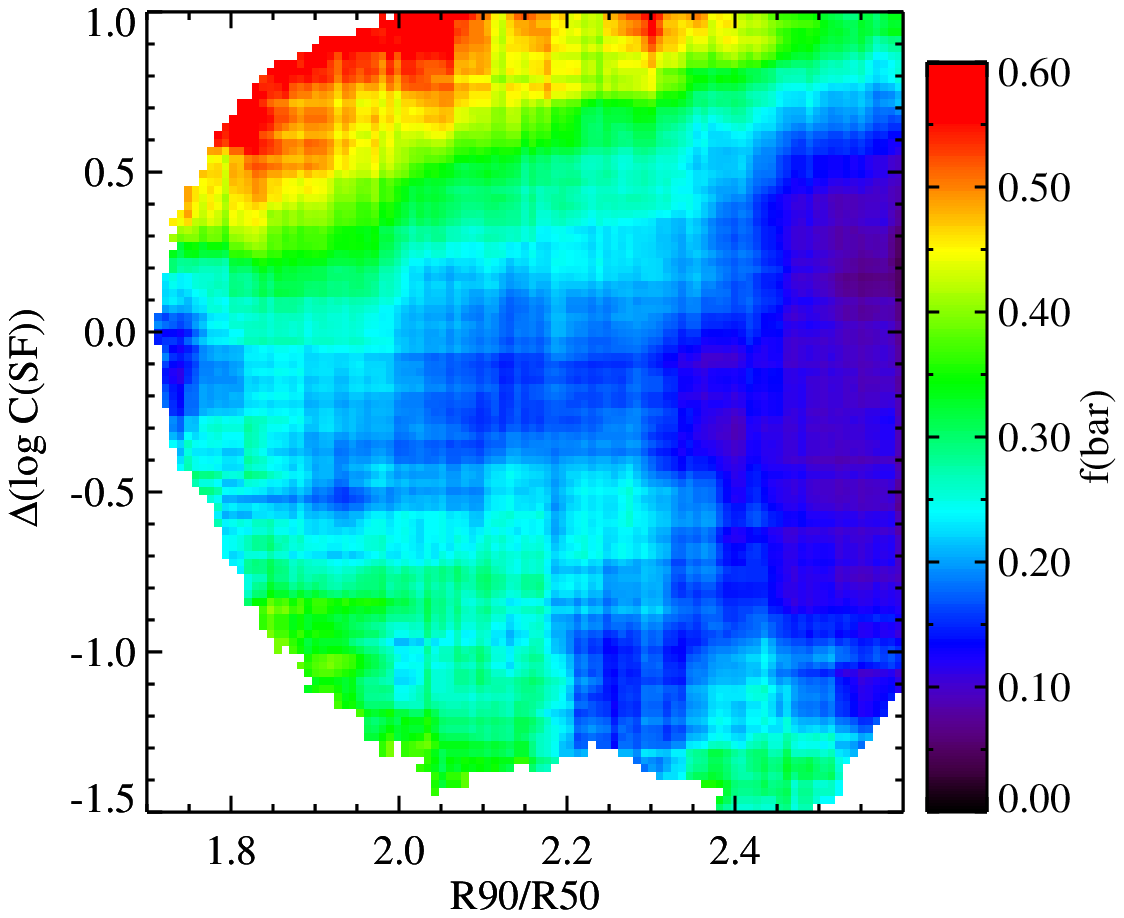}
\caption{The distribution of galaxies from Bar Sample A is plotted 
in the 2-simensional plane of of $\Delta$ C(SF) versus stellar 
mass, $\Delta$ C(SF) versus mass surface density and $\Delta$ C(SF) versus concentration. The coloured contours denote the strong bar fraction as a function of position in the plane.
}
 \label{fig:strbarfrac_sfenhance}
\end{figure*}

Finally, in Figure~\ref{fig:strbarfrac_sfenhance} we plot contours of
strong bar fraction in the plane of C(SF)   versus
stellar mass, stellar surface mass density, and concentration
parameter. This plot should tell us whether       
strong bars are not only a sufficient condition for enhanced central star formation
in galaxies, but whether they are also {\em necessary} condition. 
Note that in this plot, $\Delta(\log$ C(SF)) is defined as 
$\log$ C(SF) minus the average value for all galaxies with similar stellar mass
mass surface density and concentration from  the ``parent sample'': positive values indicate
that star formation is more concentrated than average, negative
values that it is less concentrated. 

We  see from the top row of the figure that the strong bar fraction 
peaks in galaxies with
the most concentrated star formation and with stellar masses greater than
$\sim 3 \times 10^{10} M_{\odot}$, stellar surface densities less than
$\sim 3 \times  10^8 M_{\odot}$ kpc$^{-2}$, and concentration parameters
less than $\sim 2.4$.  
The peak strong bar fraction is around 0.5.  
This implies that although bar-driven
inflows constitute  a major channel for inducing enhanced central star formation
in disk-dominated galaxies, they are not the only process at work. As shown by
\citet{Li08, Reichard09}, galaxies with close companions and lopsided galaxies
also have enhanced star formation rates. We note that in  the highest surface density
galaxies, bars appear to play no role at all in inducing enhanced central
star formation.  

Very interestingly,  there appears to be a secondary peak in the fraction of
strong bars in galaxies with the same stellar masses, stellar
surface densities and concentrations, but with central star formation
rates that are significantly {\em below average}. 
In other words, there also appears to be a population of {\em quenched galaxies}
with strong bars. This is consistent with the finding in \citet{Sheth05} 
that some barred spirals have nuclear regions that are deficient in gas.

This result raises the question as to whether 
central starbursts produced by bar-driven inflows
play any role in  shutting down star formation in some galaxies, either by
using up the available gas more quickly, or by generating winds/outflows 
that expel the gas from the central regions of these systems. 
We find no enhancement of optical or radio AGN fraction in  this population, which
suggests that starburst-driven processes may be at work.  Another possible 
explanation is that the gas has simply been consumed and that the 
galaxies are in temporarily quiescent state.
This will be the subject of a future investigation.

Finally, we note that for the low density, low concentration galaxy population 
($\mu_* < 3 \times 10^8 M_{\odot}$ kpc$^{-2}$, $C<2.5$), 
the strong bar fraction is smallest when
there is no enhancement in central star formation rate. This is consistent with
the idea that in the absence of any dynamical perturbations,
the star formation rates in present-day disk galaxies are regulated
by accretion from the surrounding halo (e.g. Kauffmann et al 1993).
Bar-driven inflows then play a key role in regulating the rate at which the       
accreted gas in consumed into stars.

\subsection{The relation between bar properties and global galaxy properties}

\begin{figure*}
 \centering
\includegraphics[width=7cm]{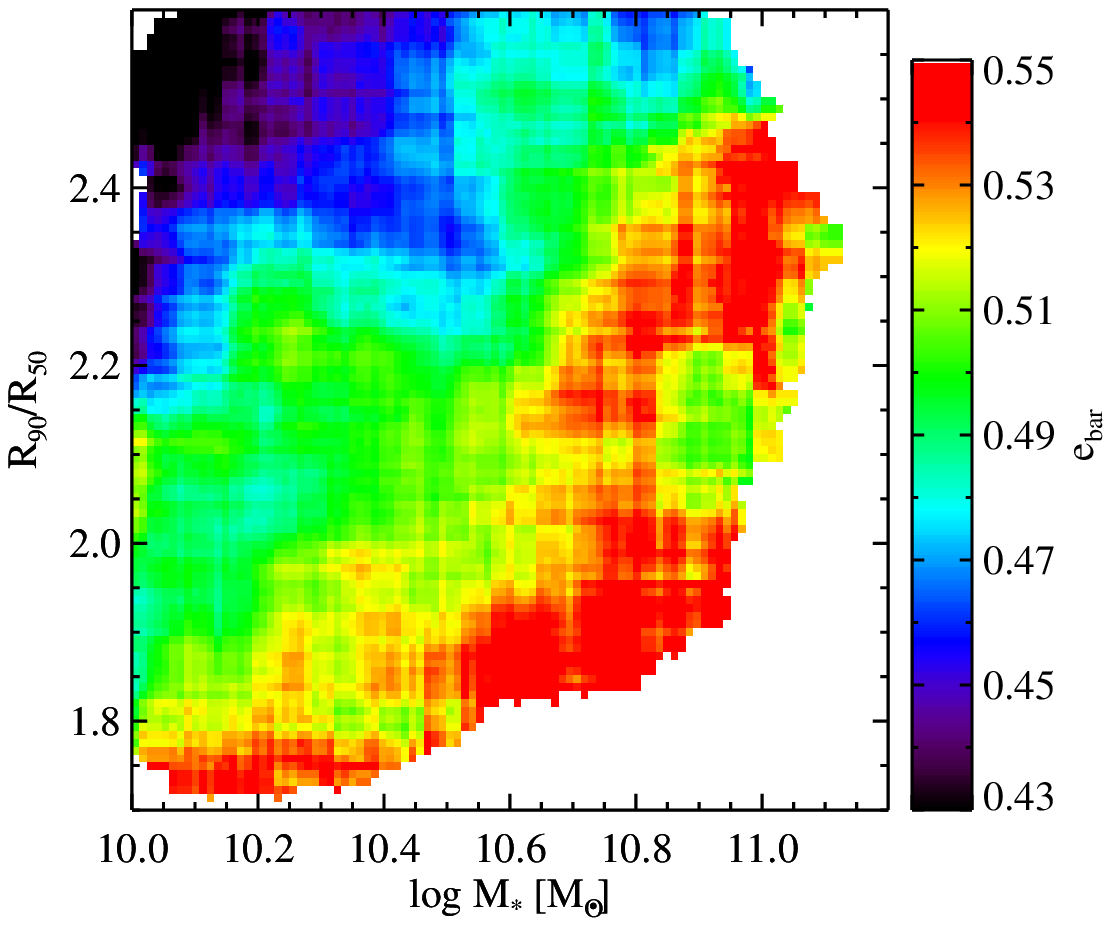}
\includegraphics[width=7cm]{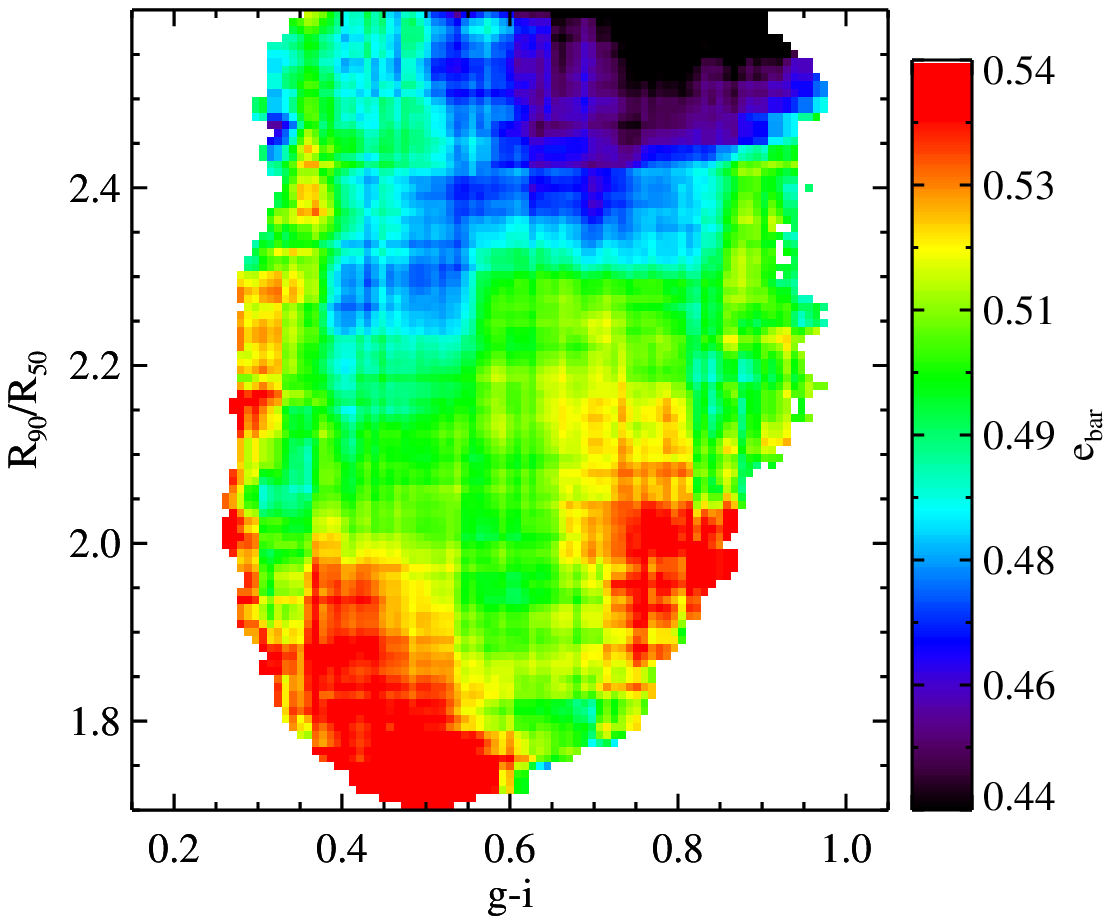}
\includegraphics[width=7cm]{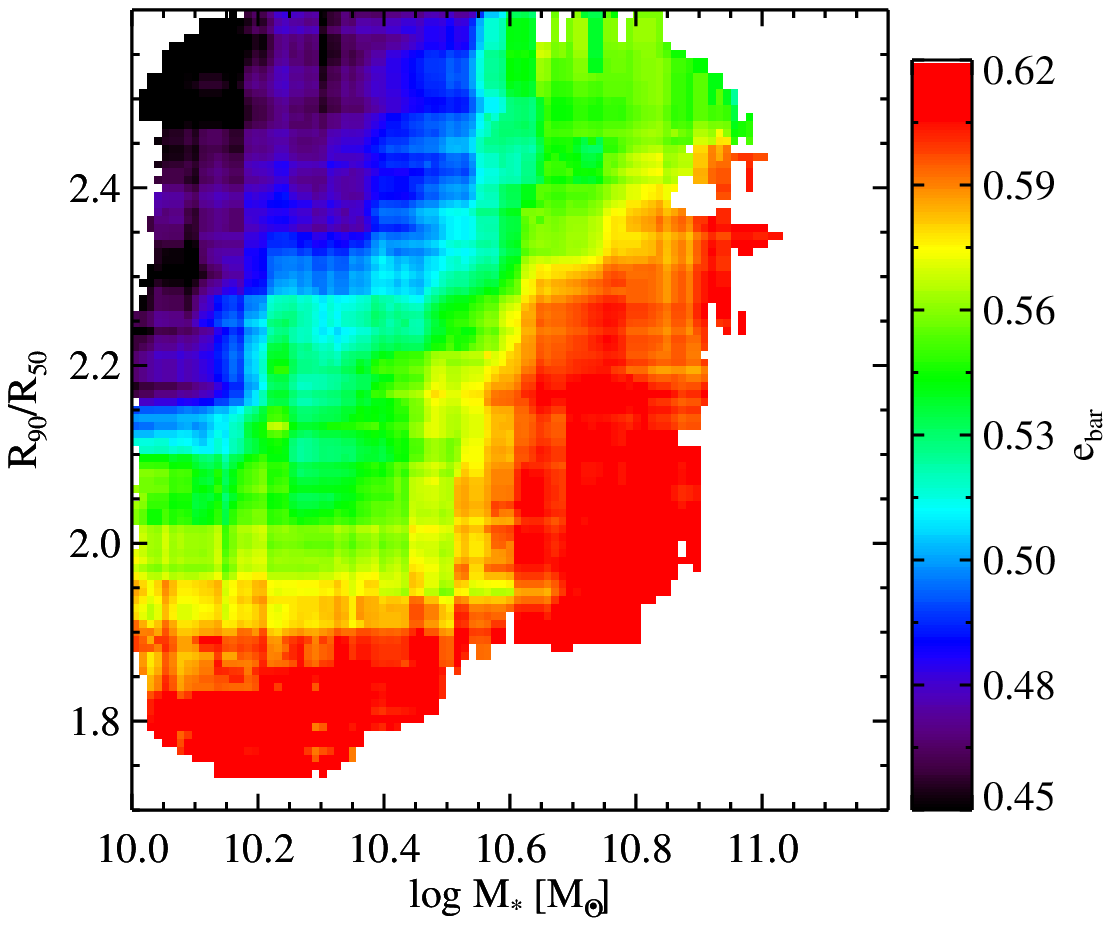}
\includegraphics[width=7cm]{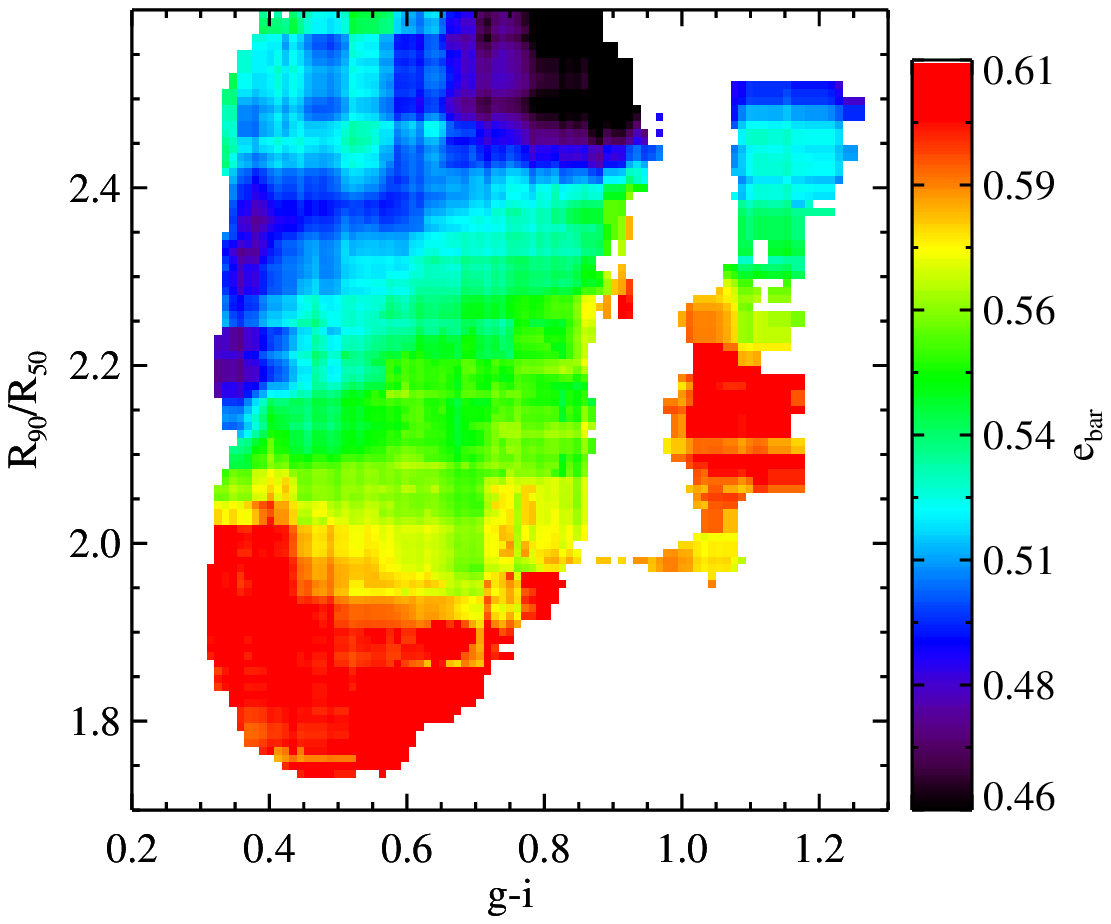}
\caption{Galaxies from Bar Sample A (top row) and 
Bar Sample B (bottom row) are plotted in the 2-dimensional plane 
of concentration versus stellar mass and concentration versus $g-i$ colour. The coloured contours denote the average value of $e_{bar}$ as a function of position in the plane.}
 \label{fig:barstr}
\end{figure*}

\begin{figure*}
\centering
\includegraphics[width=7cm]{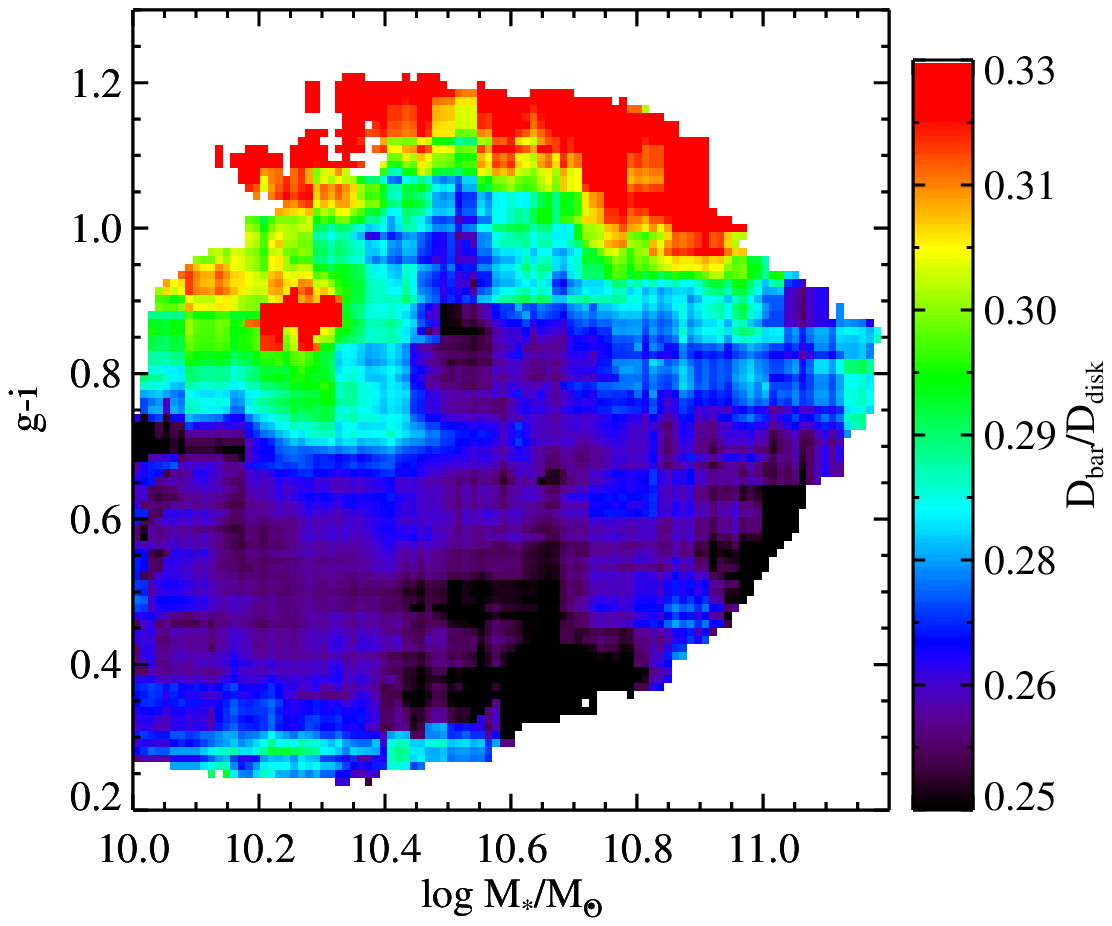}
\includegraphics[width=7cm]{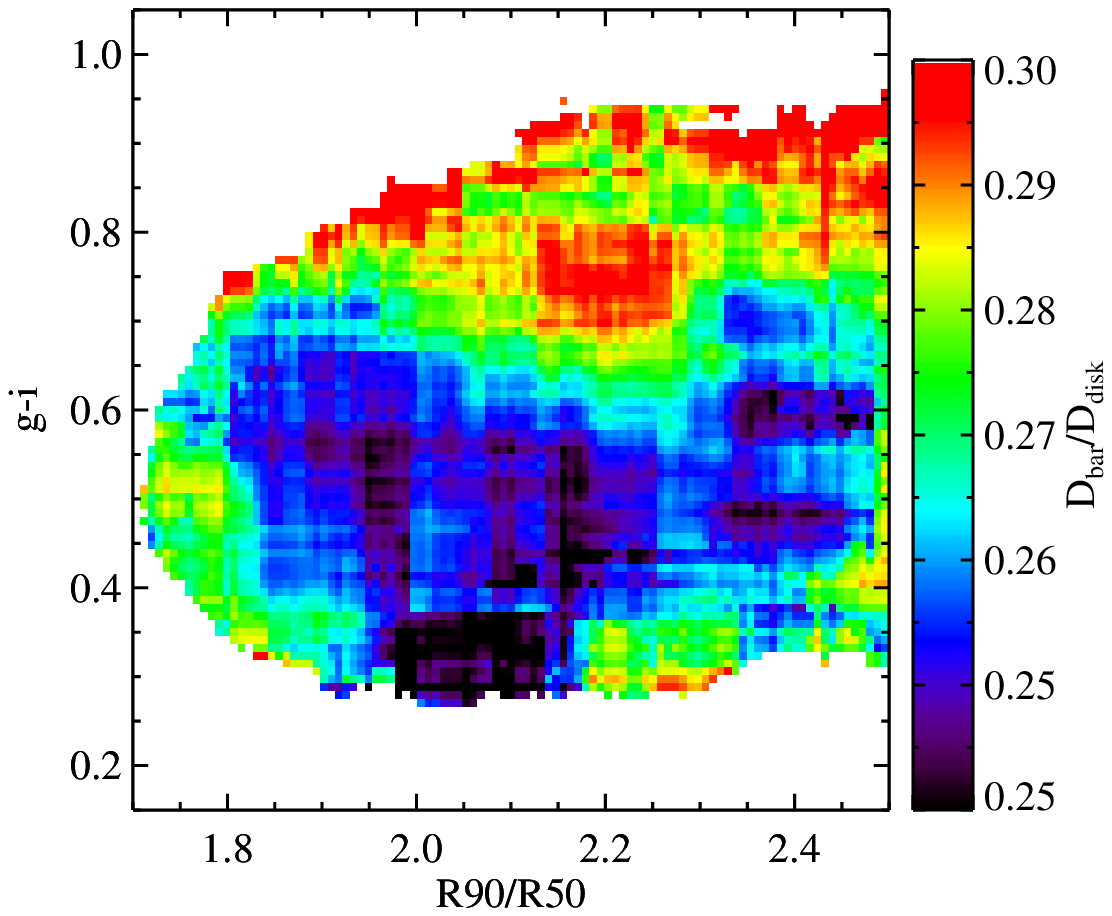}
\caption{Galaxies from Bar Sample A  are plotted in the
2-dimensional  plane of stellar mass and concentration versus $g-i$ colour. 
The coloured contours denote the average value of $D_{bar}/D_{disk}$ 
as a function of position in the plane.}
 \label{fig:barlen}
\end{figure*}

In Figure ~\ref{fig:barstr} and  Figure~\ref{fig:barlen}, we examine how 
 $e_{bar}$ and $D_{bar}/D_{disk}$ depend on global galaxy properties
such as stellar mass, concentration index and  $g-i$ colour. 

The left panels of Figure~\ref{fig:barstr} show that more massive 
and less concentrated galaxes tend to have stronger bars. This
is in agreement with results discussed previously, showing that the 
fraction of galaxies with strong bars peaks for massive, disk-dominated galaxies (see Appendix for a discussion about bulge contamination).  
The right panel shows that at a fixed concentration,
there does not appear to be any direct connection between $e_{bar}$ and
global  $g-i$ colour. Once again, results for Bar Sample A and B are  
qualitatively very similar.

In Figure~\ref{fig:barlen}, we again see that $D_{bar}/D_{disk}$ is strongly 
correlated with global $g-i$ colour. At a fixed stellar mass or concentration, 
redder galaxies have higher bar-to-disk size ratios.  In some extreme cases,  
the whole galaxy consists of a long bar and a surrounding ring.
These galaxies are almost all red. Figure 12 shows a compendium of
SDSS images of such galaxies. It is interesting to note that 
at a fixed colour, $D_{bar}/D_{disk}$ does not vary as 
a function of concentration or stellar mass. 

In summary, $e_{bar}$ is correlated with both stellar mass and structural 
parameters such as concentration, but not with global colour. 
$D_{bar}/D_{disk}$ is very strongly correlated with
global colour, and only weakly correlated with mass and structural parameters. 

\begin{figure*}
\centering
\includegraphics[width=15cm]{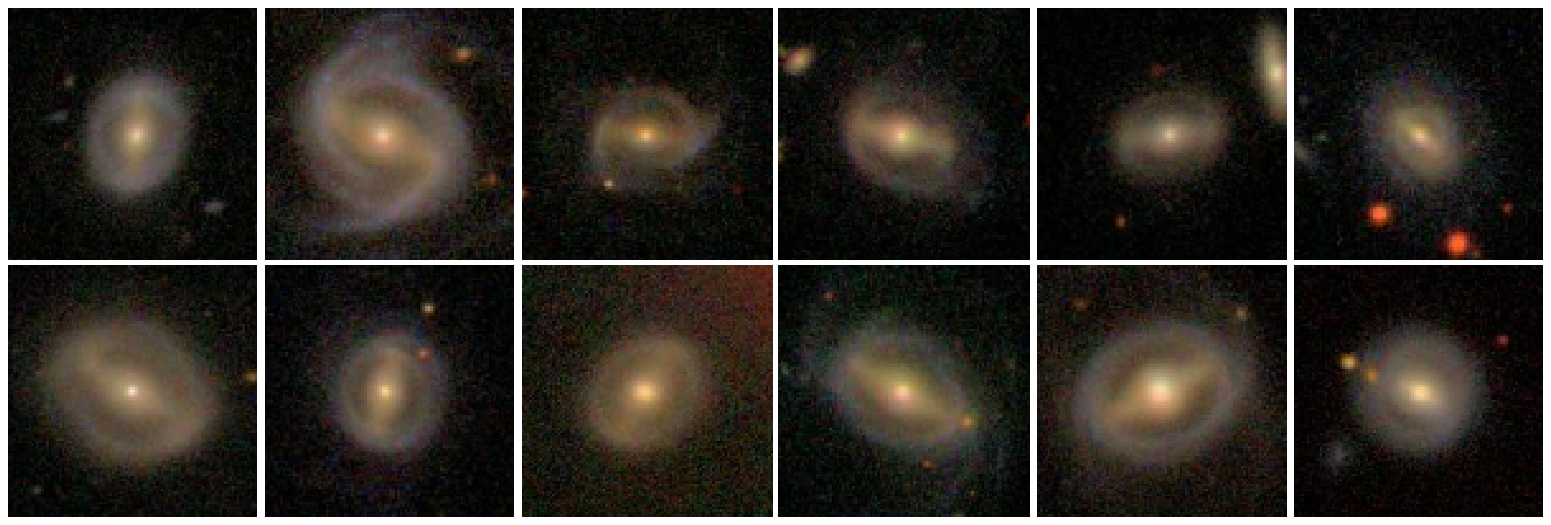}
\caption{A gallery of images of galaxies with long bars.}
 \label{fig:largebar}
\end{figure*}

\section{Summary and Discussion}

We have identified 1555 galaxies with bars longer than 5 kpc in a sample of 3757 face-on disk
galaxies with stellar masses greater than $10^{10} M_{\odot}$ and redshifts
in the range $0.01<z<0.05$ drawn from the seventh data release of the Sloan Digital 
Digital Sky Survey. We have measured the ellipticity, length and $g-i$ colour of
the bars and have explored the relationship between global
galaxy properties,  bar properties, and enhanced
central star formation  in galaxies. 

Our main conclusions may be summarized as follows:

\begin{enumerate}
\item Only strong bars with ellipticity greater than 0.5 result in  
enhanced central star formation in galaxies. This result holds for all
galaxies, regardless of their stellar mass, stellar surface mass density 
or bulge-to-disk ratio. (Section 3.1, Figure~\ref{fig:bar_ssfrcons})

\item Strong bars tend to have blue colours, and are surrounded by a 
blue outer ring. This indicates that the enhanced star formation is 
occurring on the scale of the bar itself. (Section 3.2, Figure~\ref{fig:bar_nonbar_dcl_barstr})

\item The incidence of strong bars is highest for massive galaxies
with low surface mass densities and concentrations, reaching values as high as 60\%
for galaxies with stellar masses greater than $3 \times 10^{10} M_{\odot}$,
stellar surface densities less than $3 \times 10^{8} M_{\odot}$ $kpc^{-2}$and
concentration indices less than $\sim 2.5$.  (Section 3.3, Figure~\ref{fig:strbarfrac})

\item Within this population, the incidence of strong bars appears to be bimodal.
The fraction of bars is highest for galaxies with central star formation
rates that are factor of 2-3 higher than the average. There is a secondary peak
in strong bar fraction for galaxies with central star formation rates that 
are more than a factor of 10 lower than average. (Section 3.3, Figure~\ref{fig:strbarfrac_sfenhance})

\item There is no correlation between the ellipticity (or strength) of the
bar and the global colour of the galaxy. There is a strong
correlation between the size of the bar and the global colour of the galaxy. 

\end {enumerate}

\subsection{Relation of this work to other recent studies of bars in 
large galaxy samples}

There have been a number of studies focusing on how 
the fraction and properties of  bars 
varies as a function of  Hubble type, or parameters 
such as stellar mass. Because most galaxy properties
are strongly correlated, it is difficult to identify 
which  is the {\em primary parameter} in many of these correlations.
Our approach of analyzing the properties of bars in    
two-dimensional projections of different galaxy parameters helps us
differentiate primary from secondary correlations. 

 \citet{Marinova09} (M09) found 
an increasing bar fraction as a function of galaxy
luminosity and stellar mass, consistent with our results. 
\citet{Barazza08} (B08) found the opposite  trend. 
However, our trends in bar fraction as a function of
concentration index are consistent with B08. but at odds with
 M09. The main reasons for these discrepancies are the
very different sample definitions   \citep[see discussion in][]{Nair10}. 
We note that the analysis in this paper has been carried out for a volume-limited
sample of galaxies with stellar masses greater than $10^{10} M_{\odot}$,
so the statistics presented in this paper should not suffer from
any sample selection biases.

\citet{Laurikainen07} found that $e_{bar}$ does not vary
as a function of Hubble type, \citet{Whyte02} and \citet{Aguerri09} found a 
tendency for later-type galaxies to have higher $e_{bar}$, while \citet{Martin95} found a tendency
for early-type galaxies to have higher $e_{bar}$.  Our result show a possible explanation for these discrepancy from different studies: early-type galaxies are both more 
massive and more concentrated than late-type galaxies, while we show that the trend of $e_{bar}$ varying with stellar mass is contrary to the trend with concentration.
We have chosen to bypass Hubble type altogether, focusing instead
on conecntration index and stellar surface densities as parameters
that have a simpler physical interpretation.

There have been many attempts to link the length of the bar length 
with the global structure of galaxies. Longer bars were 
found to correlate with more prominent bulges in early-type galaxies
\citep{Aguerri09,Ath80,Ath03,Erwin05,Martin95}.  
\citet{Hoyle11} was the first to find that  
galaxies hosting longer bars  also tend to be redder. Our results show that  
$D_{bar}/D_{disk}$ has the tightest correlation with 
colour and at a fixed colour $D_{bar}/D_{disk}$ does not 
vary with stellar mass or concentration. 

There have also been numerous studies of the relation between
bars,  gas concentration, 
central star formation rate and metallicity gradients in galaxies,
supporting the picture that bars induce gas inflows in galaxies  
\citep{deJong84, Jogee05, Martin95, Martin94, Sakamoto99, Sheth05, Zaritsky94}. 
Recently, \citet{Ellison11} used a large, visually classified sample
to show that barred galaxies with stellar $M_*>10^{10}M_{\odot}$
have centrally enhanced star formation rates. This is consistent with our results. 
No dependence of  central star formation enhancement on bar ellipticity
was found  in \citet{Ellison11}. One possible reason for this discrepancy is that their control sample
was  only matched in stellar mass, while our control sample was matched 
in stellar mass, surface mass density and colour.  Because the central star formation
rates of galaxies depend on all these parameters (and on colour in particular),
the  \citet{Ellison11} might not have had the sensitivity to isolate
the ellipticity dependence.
 
Finally, we note that our results are inconsistent with  some past findings 
 \citep{Devereux87,Ho97}  that late type galaxies do not have significantly enhanced central star formation .
One possible reason for this discrepancy is that these 
studies did not distinguish strongly barred galaxies 
from weakly barred galaxies. As we have seen, the fraction of
galaxies with strong bars peaks at higher stellar masses, so
the late-type samples investigated in these older studies may
have only contained galaxies with weak bars.  
We also caution that bar classifications in bulge-dominated galaxies are subject
to strong biases. Two dimensional image decomposition techniques that
allow the bulge component to be properly subtracted 
are one obvious way forward  \citep[for example,][]{Gadotti09, Weinzirl09},  but this lies well beyond the scope of this study.

\subsection {Comments on the results in this paper} 

\subsubsection {Only bars with $e_{bar}>0.5$ induce enhanced 
central star formation in galaxies}

We found that only bars with $e_{bar}>0.5$ were able to induce
enhanced levels of star formation in the central and bar-dominated regions of
the galaxies in our sample. In addition, star formation in the circular region 
surrounding the bar was also found to be enhanced. 

These results are {\em qualitatively} consistent with simulations  that 
show that gas inflows are stronger in galaxies with strong bars,
because such bars induce stronger torques that pull the gas off their
initially circular orbits, causing it to flow towards the center of the galaxy.
Some of the inflowing  gas will  stop
outside the corotation radius and form a ring \citep{Ath92, Friedli93}. 

Theoretically, the torques induced by the  bar are   proportional to 
the mass and to the elongation of the bar, but inversely proportional to the 
central axisymmetric mass \citep{Combes81a}. Observationally, these three 
factors correspond to the mass of the bar relative to that of the disk, the bar ellipticity, 
and the bulge to disk ratio of the galaxy. In order to determine the mass
of the bar, we would need to subtract a model for the bulge and the disk of the
galaxy, which is beyond the scope of this work. In this study,  
ellipticity has been used as our measure of bar strength, because
it is the more easily accessible observational parameter 
\citep{Block01, Laurikainen02a, Martinez06}. 

As the bulge-to-disk ratio increases, the effect of the bar will become weaker 
\citep{Laurikainen07, Whyte02}. Parameterizing bar strength in terms of $e_{bar}$ for
all galaxies, independent of their bulge-to-disk ratio, does not
take this effect into account. However, as we have discussed, $e_{bar}$
will be underestimated by our ellipse-fitting method in bulge-dominated 
galaxies. These two effects may somehow conspire to make $e_{bar}$
an excellent predictor of central star formation enhancement, but
we admit that further simulation studies are required to make a more
detailed interpretation of our results

\subsubsection {The fraction of strong bars in galaxies with central starbursts
is peaked at value of $0.5-0.6$ }

Tidal interactions are another obvious  mechanism for inducing enhanced 
central star formation in galaxies \citep{Cox08, Bournaud07}. 
\citet{Li08} found that $>\sim40\%$ of SDSS  galaxies with the highest central
specific star formation rates estimated have a close companion 
within a projected radius of 100 kpc). 
\citet{Li09} showed that barred galaxies 
do not have an excess of close companions \citet[similar
results were also found by also found by][]{Marinova09, Barazza09}. 
In our study, we have excluded interacting and merging galaxies 
from the ``parent sample'' in order to ensure that the ellipse
fitting procedure delivers robust results. 

In summary, our main  conclusion seems to be that bars
and galaxy-galaxy interactions are independent phenomena
that {\em together}  could explain most of the central starbursts
in galaxies in the local Universe.

\subsubsection{The size of the bar is strongly correlated with the global colour
of the host galaxy}

Simulation results indicate that once a bar is formed, it will
rapidly slow down and grow longer as angular momentum is transfered 
to the bulge and to the halo. On the other hand, if gas continues
to accrete onto the galaxy, the gas may transfer angular momentum 
to the bar, thereby re-accelerating it. This suppresses
that growth of the bar and can even cause it to become shorter  
\citep{Ath02, Ath03, Berentzen07, Curir07,  Debattista00, 
Friedli93, Villa10, Weinberg85}. Given that there is a tight correlation 
between colour and HI gas fraction in galaxies \citep{Catinella10}, 
our results suggest that the observation effect may come about
because the size growth of the bar is being regulated by gas
accretion in galaxies. Red galaxies have not had a gas accretion event for
many Gyr, so their bars have been able to grow. The fact that
the colour of the bar and the size of the bar do not correlate, 
may imply that size of the bar depends on the gas accretion rate
integrated over the past history of the galaxy, whereas the colour of the bar
depends on the {\em current} gas inflow rate.

\subsubsection { The incidence of bars peaks in disk-dominated galaxies
with high masses and low stellar surface densities}

Galaxy formation models in a $\Lambda$CDM cosmology predict that 
disk instabilities represents the dominant contribution to the formation of bulges in massive galaxies with $M_*$ between $10^{10}$ and $10^{11}$ $M_{\odot}$ \citep{DeLucia11}. 
In these studies, bars are assumed to form when
disks satisfy the standard Toomre Q-parameter criterion, which
has been validated by N-body simulations
\citep{eln1982}:
\begin{equation} \frac{V_{\rm max}}{(GM_{\rm disk}/R_d)^{1/2}} <
\epsilon\sim0.5-1.  \end{equation}

We find that $f(bar)$ peaks when $M_*$ is high, in agreement
with the \citep{DeLucia11} results. However, at fixed stellar mass,
f(bar) increases at {\em lower}  stellar lower surface densities, 
When $\mu>10^{8.5} M_{\odot}$ $kpc^{-2}$ 
or $R_{90}/R_{50}>2.6$, f(bar) quickly drops to values near 0. This
is not easily understood in the context of the formula given above. Athanassoula (2008) has criticized the \citet{eln1982} criterion, pointing out
that it is not of general applicability and should not be used in semi-analytic
models. In particular, stellar disks containing gas would be expected
to be significantly more unstable than predicted.

Studies have shown that $\mu_*\sim10^{8.5} M_{\odot}$ $ kpc^{-2}$ is a special transition 
point for galaxies in the local Universe. At surface densities lower than this value,
galaxies contain significant amounts of gas, but at higher surface densities,
the atomic and molecular gas fraction in galaxies 
drop suddenly in many galaxies \citep{Catinella10, Saintonge11}.
A similar transition is seen at $R_{90}/R_{50}\sim2.6$.
In contrast, the correlation between gas content and stellar mass 
is not as pronounced. It is thus reasonable to 
suppose that the fraction of bars may peak at stellar surface densities below $10^{8.5} M_{\odot}$ $kpc^{-2}$  
in the local Universe, because this {\em currently} represents some special 
instability point for galactic disks. What we are not able to say from these          
observations, is what role bars have played over the whole history
of the Universe in shaping this instability point.    
In any case, our results do suggest that gas accretion, 
 disk instabilities, structural changes, and quenching of gas
and star formation may happen coherently in galaxies. In the present-day Universe, bars
are clearly an important regulatory mechanism  in the ``transition regime''
between the blue and the red populations. 

\section*{Acknowledgements}
We thank A. Cooper, T. Heckman, S. Jogee, and S. White et al. for useful discussions. 

XK is supported by the National Natural Science Foundation of China (NSFC, Nos. 10633020, and 10873012), the Knowledge Innovation Program of the Chinese Academy of Sciences (No. KJCX2-YW-T05), and National Basic Research Program of China (973 Program; No. 2007CB815404).

GALEX (Galaxy Evolution Explorer) is a NASA Small Explorer, launched in April 2003, developed in cooperation with the Centre National d'Etudes Spatiales of France and the Korean Ministry of Science and Technology. 

Funding for the SDSS and SDSS-II has been provided by the Alfred P. Sloan Foundation, the Participating Institutions, the National Science Foundation, the U.S. Department of Energy, the National Aeronautics and Space Administration, the Japanese Monbukagakusho, the Max Planck Society, and the Higher Education Funding Council for England. The SDSS Web Site is http://www.sdss.org/.

\bibliographystyle{mn2e}

\appendix

\section{ Tests of our bar identification procedure}

In Sect 2.2, we discussed that  
the existence of prominent bulges will make the elliptical isophotes
of galactic bars rounder, causing us to miss bars in bulge-dominated galaxies
and under-estimate the bar ellipticity. Here we quantify 
the affect of bulge contamination on our estimates of ellipticity
and bar fraction by simulating the effect of the bulge of our measurements .

First, we select the galaxies with $R_{90}/R_{50}<2.2$ from the original barred sample. 
According to the formula from \citet{Gadotti09}, $R_{90}/R_{50}<2.2$ 
corresponds to a bulge mass fraction of less than 14$\%$. 
We use the images of these galaxies as ``inputs'', and we
then add idealized bulge components onto them as described below. 
>From now on, the bar parameters related to the original images will be referred to as
{\em intrinsic} and the  parameters related to the simulated images 
will be referred to as the {\em output}. 

A pseudo bulge is more compact than the classical bulge \citep{Gadotti09},
and is thus is more likely to distort the inner isophote of galaxies, 
so to be conservative, we only simulate pseudo-bulges in this test. 
We set the stellar mass of the pseudo bulges to range from 0.03 to 1.8 
times the stellar mass of the original galaxies. This range is chosen 
to produce the $R_{90}/R_{50}$ range of the galaxies in our  ``parent sample''. 

>From Figure 13 of \citet{Gadotti09}, we estimate 
$\log r_e\sim0.2\times$ log $M_*-2.05$, where $r_e$  and $M_*$ 
are the effective radius and stellar mass of the pseudo-bulge. 
We describe the surface brightness of a pseudo-bulge by a sersic 
profile with index 1.4, $I(r)=I_0\times exp(-2.47\times(r/r_e)^{1/1.4})$, 
where $I(r)$ is the surface brightness at radius $r$, and $I_0$
is the central surface brightness  \citep{KK04}.
We assume that the $i$-band  flux  is linearly proportional 
to the stellar mass \citep{Bell03}. Integrating  the surface 
brightness profile to get the total flux, $I_0$ is estimated 
as $I_0=flux/(1.17\times r_e^2)$. 

\begin{figure*}
\centering
\includegraphics[width=14cm]{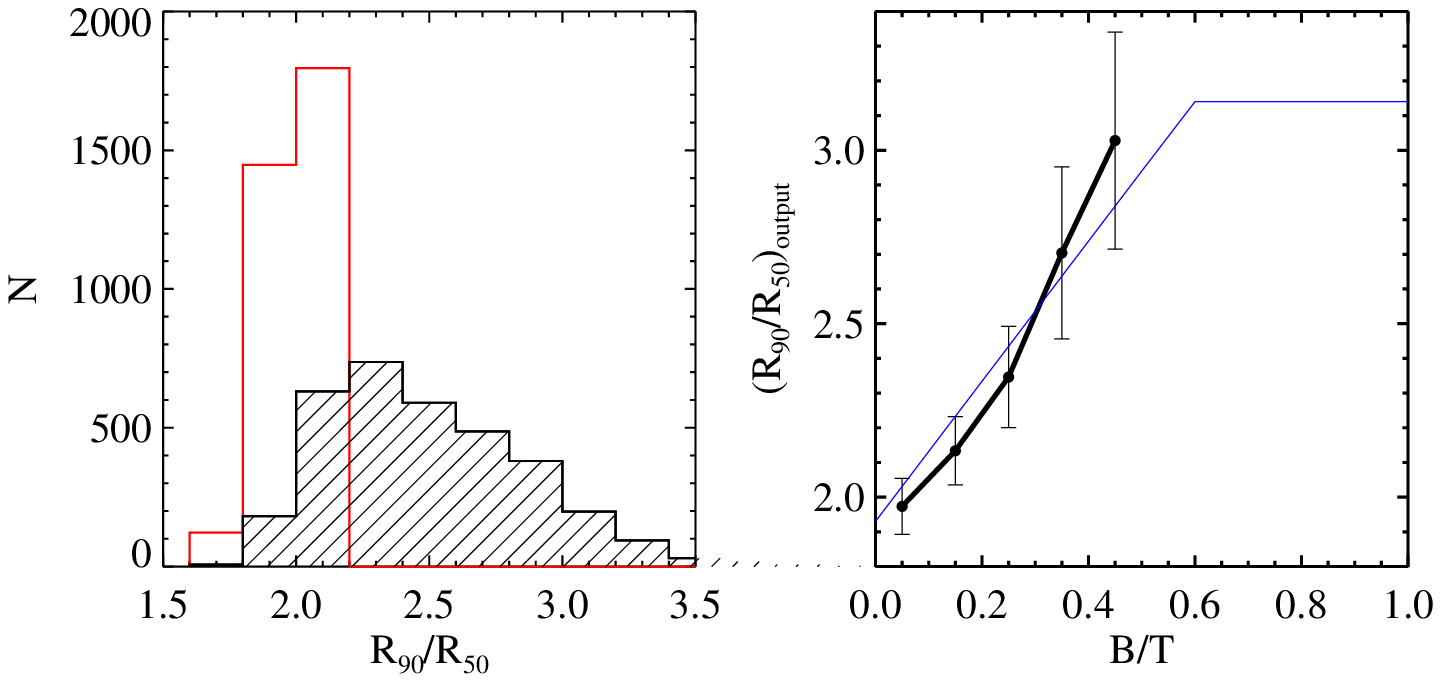}
\caption{The left panel plots the distribution of input (red histogram) 
and output (black, hatched histogram) $R_{90}/R_{50}$ values. In the right panel,
the black line shows the correlation between output concentration 
and the bulge-to-disk ratio. The blue lines shows  
the relation between concentration and bulge-to-disk ratio from
the 2-dimensional decompositions of  \citet{Gadotti09}.}
 \label{fig:cons_distr}
\end{figure*}

We  produce 2000 simulated galaxy images by adding the 
simulated pseudo-bulge components to the real $i$-band images 
of the disk-dominated galaxies. The distribution of  concentration 
indices ($R_{90}/R_{50}$) of the simulated galaxies is shown 
in Figure~\ref{fig:cons_distr}. We can see that the 
simulated $R_{90}/R_{50}$ ranges from 1.8 to $\sim$3, and follows 
the same relation with bulge-to-total mass ratio of the 
galaxy shown in \citet{Gadotti09}. 

We then use the ellipse-fitting method described in Sect. 2.2 to 
identify and measure  bars from the simulated images. We refer to 
the fraction of bars identified from the simulated images 
as the output bar fraction ($F_{output}(bar)$), and the correspondingly 
measured bar ellipticity as $e_{bar,output}$.

\begin{figure*}
\centering
\includegraphics[width=5cm]{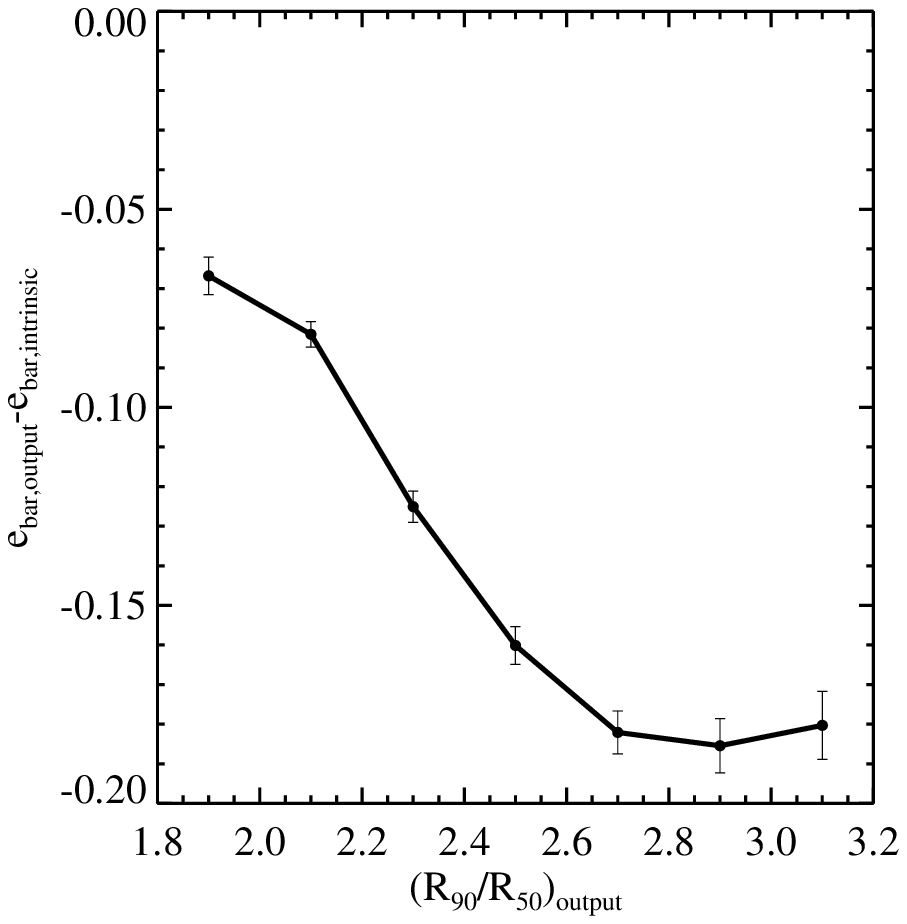}
\includegraphics[width=5cm]{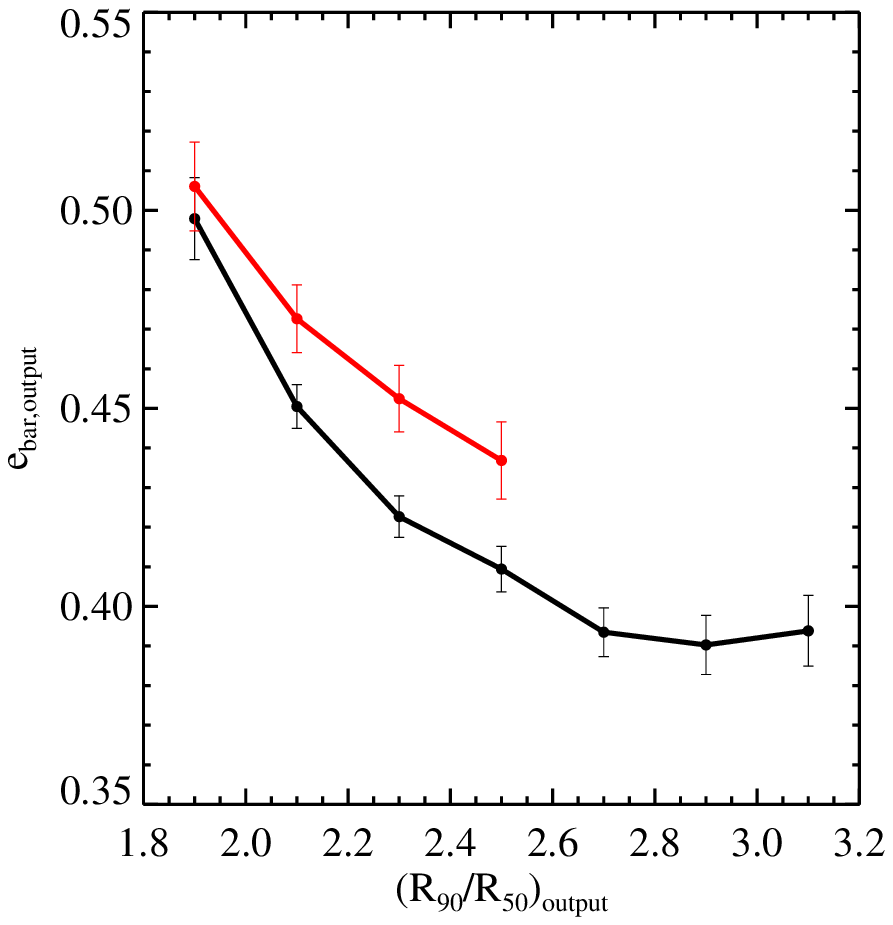}
\includegraphics[width=5cm]{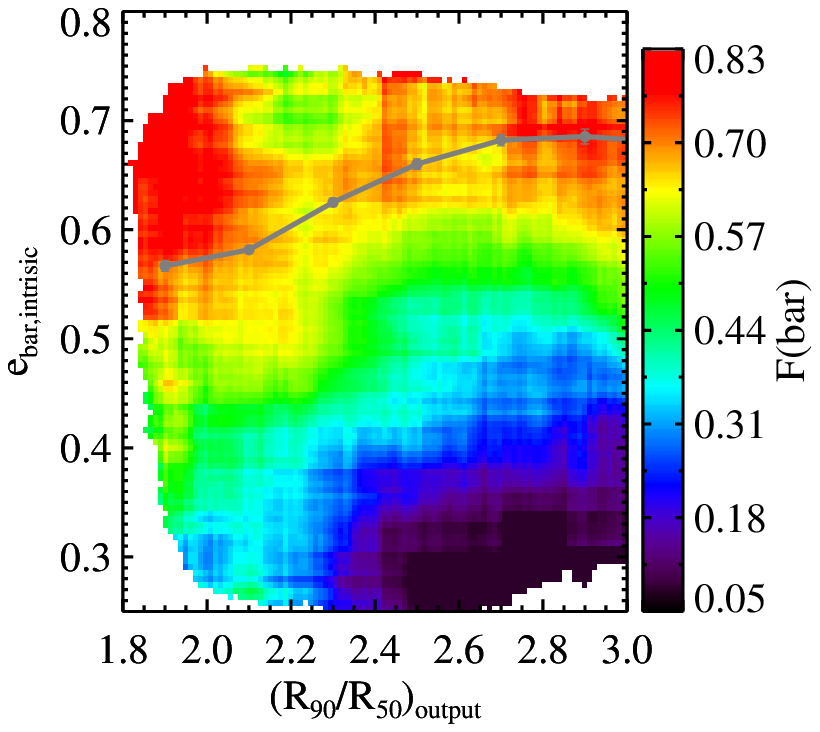}
\caption{In the left panel, ($e_{bar,output}-e_{bar,intrinsic}$) 
is plotted as a function of $R_{90}/R_{50}$. The line shows the average value,
with the error bars calculated using  bootstrapping. In the middle panel, $e_{bar,output}$ (black line) and $e_{bar,obs}$ (red line)
from real galaxies are plotted as a funciton of $R_{90}/R_{50}$. In the right panel, 
the distribution of simulated galaxies is plotted in the
2-dimensional  plane  of output concentration $R_{90}/R_{50}$ 
versus input $e_{bar,intrinsic}$. The coloured contours denote 
the bar verification rate ($F(bar)$) as a function of 
position in the plane. The grey curve  shows the average 
input $e_{bar,intrinsic}$ that corresponds to an $e_{bar,outout}$
value of 0.5, as a function of $R_{90}/R_{50}$.}
 \label{fig:bar_ellipse_b2d}
\end{figure*}

In the left panel of Figure~\ref{fig:bar_ellipse_b2d}, we show how 
the intrinsic ellipticity of the bar ($e_{bar,intrinsic}$) is 
 under-estimated as the concentration increases.  We see 
that when  $R_{90}/R_{50} =$1.8, 2.4 and 2.8, $e_{bar,intrinsic}$ is 
under-estimated by 0.065, 0.15 and 0.18. When $R_{90}/R_{50}>2.7$, 
the error in flattens at a value of $\sim$0.18. 

In the middle panel of Figure~\ref{fig:bar_ellipse_b2d}, we show how the $e_{bar,output}$ from simulation 
and the $e_{bar,obs}$ measured from real galaxies vary as a function
of concentration $R_{90}/R_{50}$. The relation between  $e_{bar,output}$ and $R_{90}/R_{50}$ (black line) shows the pure bulge contamination effect.
We can see that the bulge contamination could be responsible for all the decreasing trend of $e_{bar}$ as a function of 
$R_{90}/R_{50}$. Thus it is unclear whether there is a intrinsic trend that when $R_{90}/R_{50}$ is lower, $e_{bar,intrinsic}$
is higher.

In the right panel of Figure~\ref{fig:bar_ellipse_b2d}, we show the 
bar verification rate ($F(bar)$) from the simulated images 
in the plain of output concentration $R_{90}/R_{50}$ versus input $e_{bar,intrinsic}$. 
We can see that at a fixed $R_{90}/R_{50}$, $F(bar)$ is strongly 
correlated with input $e_{bar,intrinsic}$. At $R_{90}/R_{50}\sim2.6$, 
when  $e_{bar,intrinsic}\sim0.3$, only $\sim20\%$ of the barred 
galaxies are identified. This is expected because
we require $e_{bar,output}>0.25$ for a galaxy to be classified
as barred.   When $e_{bar,intrinsic}\sim0.5$, $\sim50\%$ of 
the barred galaxies are identified. When $e_{bar,intinsic}>0.6$, $F(bar)$ 
flattens at a value of about 0.7.  Bulge contamination thus results
in loss of bars with low values of $e_{bar}$. 

In this study, we focus on galaxies which have $e_{bar,obs}\sim0.5$. 
This corresponds to an input $e_{bar,intrinsic}$  that is
a weakly rising function of  $R_{90}/R_{50}$, as
shown by the grey curve in the right panel of Figure~\ref{fig:bar_ellipse_b2d}). 
We see that most of the region above the $e_{bar,obs}\sim0.5$ 
curve corresponds to $F(bar)$ values greater than 70\%. 
In the region enclosed by the curve and $e_{bar,intrinsic} \sim 0.5$, 
$F(bar)$ can vary from values of around 70\% for the less concentrated galaxies to values of around 50\% for the more concentrated galaxies.
We conclude, therefore, that we are likely correctly identifying between 2/3 and 3/4
of the barred galaxies with intrinsic ellipticities greater than 0.5 

\end{document}